\documentclass[runningheads]{llncs}
\usepackage[margin=1in]{geometry}
\usepackage{graphicx}
\usepackage{subcaption}
\usepackage{paralist, tabularx}
\usepackage{amssymb}
\usepackage{mathtools}
\usepackage{amsfonts} 
\usepackage{xspace}
\usepackage{cite}
\usepackage{url}

\usepackage{enumitem}
\usepackage{multicol}
\usepackage{hyperref}
\usepackage{stmaryrd} %
\usepackage{xcolor} %
\usepackage{pdflscape} %
\usepackage{tikz}
\usetikzlibrary{positioning, arrows.meta, shapes.geometric, fit, calc, backgrounds}
\setlist[itemize]{label=-}

\usepackage{pdfpages} 
\usepackage{booktabs}
\usepackage{multirow}
\usepackage{float}
\usepackage{listings}

\title{Isolation Failure From Shared Storage: Characterizing and Exploiting Page-Cache SCA Leakage Across Containers and VMs}
\author{
Alon Abudraham\inst{1} \and
Xingyu Chen\inst{2} \and
Itamar Levi\inst{1} \and
Ari Trachtenberg\inst{2}
}

\institute{
Bar-Ilan University, Ramat Gan, Israel
\email{\{alon.abudraham,itamar.levi\}@biu.ac.il}
\and
Boston University, Boston, MA, USA
\email{\{chxy517,trachten\}@bu.edu}
}

\usepackage{algorithm}
\usepackage{algpseudocode}
\usepackage{bbm}
\usepackage{silence}
\let\subparagraph\paragraph

\newcommand{\eg}{e.g.,\xspace}
\newcommand{\topOne}{Top\nobreakdash-1\xspace}
\newcommand{\topTwo}{Top\nobreakdash-2\xspace}

\begin{document}

\maketitle

\begin{abstract}

Modern cloud platforms increasingly combine strong software isolation mechanisms with shared hardware resources to improve performance and resource efficiency. Conventional containers do this by sharing the host kernel directly, whereas sandboxed runtimes (\eg gVisor) and VM-based runtimes (\eg Kata, QEMU/KVM) provide progressively stronger isolation. In all cases, when tenants access host-backed filesystem state, the host page cache can remain shared and observable.  Although OS-managed, this page-cache channel forms an \emph{OS-mediated microarchitectural timing side channel} whose signal is shaped by the processor microarchitecture, memory and storage hierarchies, and virtualization mechanisms.  We thus investigate whether unprivileged timing measurements can reveal page-cache residency across these isolation boundaries.  Our evaluation covers Docker; gVisor with systrap and KVM; Kata Containers using QEMU and Cloud Hypervisor with shared host filesystems; Kata using QEMU, Cloud Hypervisor, and Firecracker with block-device-backed storage; and QEMU/KVM virtual machines under multiple host cache policies.  Our results show that the timing signal persists whenever the I/O path exposes shared, host-cacheable file-backed objects, including under OverlayFS layers, virtio-fs exports, and loop-backed block devices.  On the other hand, direct I/O and dedicated block devices substantially attenuate or eliminate the signal.  Virtualization therefore reshapes leakage through added latency and algorithmic noise but does not remove the underlying dependence on shared hardware and cache state. We showcase this through a case study in which we recover coarse-grained activity from a WordPress deployment backed by MySQL. These results place page-cache attacks within the broader class of OS-mediated microarchitectural timing channels and motivate coordinated hardware, virtualization, and OS support for timing isolation.


\end{abstract}

\section{Introduction}

Modern cloud platforms rely on aggressive resource sharing. Public clouds, private clusters, and research infrastructures commonly colocate mutually untrusted workloads on the same physical machines in order to improve hardware utilization and reduce operational cost. Isolation mechanisms such as Linux containers, sandboxed container runtimes, and hardware-assisted virtual machines are therefore expected to provide strong security boundaries while still allowing the underlying system to share resources efficiently~\cite{docker,kubernetes,linuxKVM,gvisor,borg,burns2016borg}.

One important shared resource is the operating-system page cache. The page cache stores recently accessed file-backed pages in memory to avoid repeated disk accesses~\cite{linuxPageCache}. This design is essential for performance, especially when many workloads access the same binaries, libraries, container image layers, or configuration files. However, the same mechanism creates shared state: whether a file page is resident in memory depends on the access patterns of other workloads on the host. Consequently, page-cache residency can encode information about victim activity and serves as a side or covert channel.

Prior work showed that page-cache state can be exploited within a single operating system and across conventional container boundaries. Gruss et al.~\cite{pageCache} demonstrated practical page-cache side channels using interfaces such as \texttt{mincore}. Boskov et al.~\cite{boskov2022union} later showed that container image sharing and OverlayFS semantics can expose similar leakage across Docker containers and Kubernetes deployments. These attacks established that shared page-cache state is dangerous, but they leave an important question open: does page-cache leakage persist when the victim and attacker are separated by stronger isolation mechanisms that mediate system calls or run workloads inside virtual machines?

In this paper, we investigate the Linux page cache as an \emph{OS-mediated microarchitectural timing side channel}. Although page-cache residency is managed by the operating system, the observable leakage is shaped by processor microarchitecture, memory hierarchy, storage subsystem, and virtualization mechanisms. 
As with conventional cache attacks such as Prime+Probe~\cite{osvik2006cache}
or Flush+Reload~\cite{yarom2014flush},
 an adversary infers shared state from latency variations, which themselves are corrupted by hardware-originating noise. 
 
We therefore view the page cache as part of the broader family of architectural and microarchitectural timing side channels rather than as a purely software artifact. The operating system exposes and manages the page cache, \emph{but the leakage fundamentally exists because of efforts to bridge hardware latency gaps} between persistent storage, main memory, and computation. Virtualization technologies —including containers, sandboxed runtimes, and virtual machines— primarily reshape the observable signal by introducing additional latency, scheduling variability, and memory-management effects rather than eliminating the underlying leakage mechanism. This perspective motivates a systematic characterization of how different hardware/software stacks influence the strength, fidelity, and exploitability of the resulting side channel. 

We evaluate conventional Docker containers, gVisor with both systrap and KVM backends, Kata Containers with QEMU, Cloud Hypervisor, and Firecracker, and QEMU/KVM virtual machines under \emph{different hardware-storage configurations}. Unlike attacks that rely on explicit page-residency interfaces, we use \emph{only unprivileged timing}: the attacker measures the latency of accessing selected file-backed pages and infers whether those pages were recently accessed by another isolated workload.

\begin{figure}[htbp]
  \centering
  \resizebox{0.80\linewidth}{!}{%
  \includegraphics[width=\linewidth]{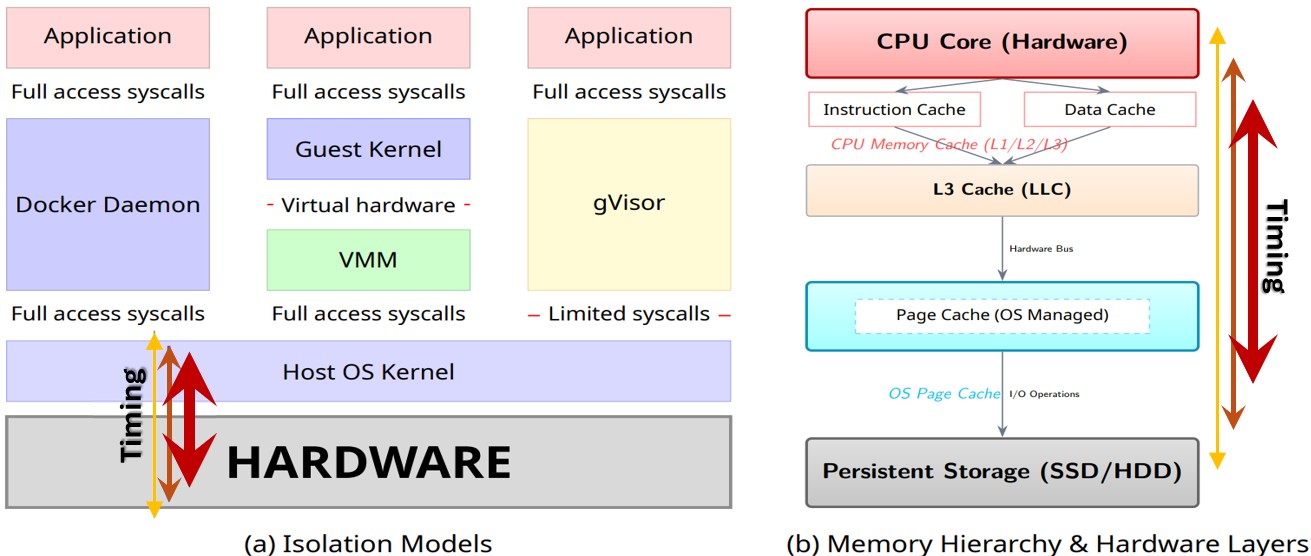}}
  \caption{Hardware-Software interfaces and Isolation boundaries considered in this work (a) Containers, sandboxed containers, and virtual machines introduce different execution boundaries, but may still interact through host-backed shared-HW storage
and page-cache state. (b) Comparison of hardware caches (L1--L3) and the software-managed page cache.}
  \label{fig:isolation_comparison}
\end{figure}

Our central finding is that the leakage characteristics of the page-cache microarchitectural side channel are jointly determined by the isolation mechanism and the underlying hardware-supported storage path. Runtime boundaries primarily modify the channel by attenuating the timing signal and increasing algorithmic and architectural noise, whereas the presence of shared host-cacheable objects determines whether exploitable microarchitectural state remains observable.
When the runtime exposes shared host-cacheable file-backed objects, as happens with OverlayFS layers, virtio-fs exports, host-backed VM images, or loop-backed block devices, cross-boundary accesses can create measurable page-cache timing shifts. In contrast, configurations based on direct I/O or real block-device-backed storage substantially weaken or remove the signal in our experiments. Thus, the security question is not simply whether a workload runs in Docker, gVisor, Kata, or QEMU/KVM, but whether the storage backend utilizes a shared host page-cache surface.

We evaluate the practical relevance of this leakage through a
database-backed web application. In a  MySQL backed WordPress deployment, we show that page-cache observations over selected MySQL pages can distinguish coarse-grained %
actions. Rather than demonstrating an %
end-to-end attack, this study seeks to demonstrate that the timing signal is strong enough to expose semantically meaningful %
behavior across isolation.%

\subsection{Contributions}

This paper makes the following major contributions:

\begin{enumerate}
  \setlength{\itemsep}{0pt}
  \setlength{\parsep}{0pt}
    \item \textbf{OS-mediated microarchitectural timing side-channel.} %
    We demonstrate that the Linux page cache constitutes an OS-mediated microarchitectural timing side channel whose leakage is governed by processor architecture, storage path, shared memory hierarchy, and virtualization mechanism, persisting
    across several modern isolation mechanisms, even when explicit page-residency interfaces
    (like \texttt{mincore}) are unavailable or uninformative.

    \item \textbf{Signal and leakage characterization.} %
    We experimentally characterize the influence of storage backends, virtualization layers, and hardware configurations on timing leakage, signal strength, and the effective Signal-to-Noise Ratio (SNR) of the page-cache microarchitectural side channel. %

    \item \textbf{Cross-isolation microarchitectural evaluation.} %
    We evaluate Docker, gVisor, Kata Containers, Firecracker, and QEMU/KVM, demonstrating that virtualization reshapes—but does not eliminate—the underlying leakage. Surprisingly, background interference from unrelated processes may even amplify the timing leakage.

    \item \textbf{Application-level side-channel case study.}
    We demonstrate that timing analysis can reveal coarse-grained
    WordPress/MySQL activity across isolation boundaries, allowing identification of several user and
    administrator actions.%

    \item \textbf{Implications for secure system architecture.} %
    We show that mitigating page-cache microarchitectural side channels requires reasoning jointly about the memory hierarchy, storage architecture, virtualization mechanisms, and operating-system policies rather than about software isolation alone. %
    
\end{enumerate}

In total, these contributions suggest that page-cache leakage should be addressed through hardware/OS co-design and stronger architectural support for timing isolation in future shared-computing platforms.

\subsection{Roadmap}
The remainder of this paper is organized as follows.
Section~\ref{sec:background} provides the necessary background on the Linux page cache, container and virtual-machine isolation mechanisms, storage paths such as OverlayFS, virtio-fs, and block-device-backed snapshots, and prior page-cache side-channel attacks. Section~\ref{sec:attack} defines our threat
model and attacker assumptions. Section~\ref{sec:eval} presents our evaluation methodology and results, including cross-boundary timing measurements, storage-backend comparisons, hardware-platform sensitivity, the effect of cache pressure, and an application-level WordPress/MySQL side channel. Section~\ref{sec:conclusions} concludes the paper.%

\section{Background \& Related Work} %
\label{sec:background}

We next provide the background needed to understand why page-cache
leakage depends jointly on the isolation mechanism, the storage path, and hardware-caching mechanisms.
We first review the Linux page cache and container image sharing.  We then examine how
sandboxed containers and virtual machines expose storage to isolated workloads, focusing on common architectures:  OverlayFS, virtio-fs, QEMU/KVM disk cache modes, and block-device
backends. Finally, we discuss prior page-cache attacks through kernel-state hardware-mediated side-channels.

\subsection{Operating-System Page Cache} %
The latency gap between DRAM memory and secondary storage is a longstanding performance bottleneck. The Linux page cache~\cite{linuxPageCache} mitigates this gap by caching filesystem-backed pages in memory. On a file \emph{read}, the kernel checks page-cache metadata: hits are served from memory, whereas misses access storage, typically at orders-of-magnitude higher latency. %
 The page cache supports both buffered and memory-mapped I/O and manages write-back and eviction. Because it is shared system-wide and keyed by the backing object rather than process identity, workloads accessing the same libraries, container image layers, or configuration files can reuse cached pages and avoid redundant I/O.

In containerized environments, union filesystems such as OverlayFS combine shared read-only lower layers with per-container, copy-on-write upper layers. Lower-layer pages remain in the host page cache and are therefore shared among containers instantiated from the same image~\cite{docker}.
Although page residency is managed by the operating system, its timing effects arise from the shared hardware memory hierarchy and storage path. The page cache therefore forms OS-mediated microarchitectural state, enabling timing channels analogous to those involving the Last Level Cache (LLC), Translation Lookaside Buffer (TLB), and Dynamic RAM (DRAM). Page-cache residency can consequently reveal a coarse-grained trace of recent file activity.

Gruss et al.~\cite{pageCache} demonstrated that this shared state is externally observable and can be turned into a mediated observable of the hardware side channel, allowing an unprivileged adversary to infer which file pages a victim has touched, from the use of hardware resources. Subsequent work and industry advisories have recognized this behavior as a distinct vulnerability class (CVE-2019-5489), affecting Linux kernels and other operating systems that expose page-cache residency through interfaces such as \texttt{mincore()} or \texttt{QueryWorkingSet()}. Our work builds on this line of research, but focuses on \emph{purely timing-based observations} of page-cache residency in the absence of explicit residency interfaces.
This object-based cache identity is the root of the leakage studied in this
paper: if two isolated workloads ultimately access the same host-backed file
object, one workload can change the timing behavior observed by the other utilizing the underlying hardware.

\subsection{Container Image Sharing and OverlayFS}
Linux containers provide process-level isolation by combining namespaces, control groups (cgroups), and security mechanisms (\eg \emph{capabilities}, \emph{seccomp}, and mandatory access control), while reusing the host kernel~\cite{docker}. Container engines like Docker and Kubernetes represent filesystem state as layered images, typically implemented using union or overlay file systems such as OverlayFS~\cite{docker,kubernetes}. In this model, a set of read-only lower layers encodes the base image, while each container receives a private upper layer that captures its modifications; the union mount presents a unified directory tree to the application, with copy-on-write semantics for updates. %
From a performance perspective, this design allows many containers to share the same read-only image layers, reducing storage overhead and improving image distribution efficiency. However, it also implies that the underlying physical file pages for the lower layers are shared in the host page cache across all containers derived from the same host-backed container image layer. Boskov et al.'s Union Buster \cite{boskov2022union} has shown that this shared page-cache state, together with overlayfs semantics, suffices to construct simple cross-container covert channels and infer some application-level behavior.%

\subsection{gVisor} 
The gVisor container sandbox~\cite{gvisor} is a Google-developed user-space kernel that
improves container isolation while efficiently sharing hardware.
Primarily deployed in the Google Kubernetes Engine (GKE) on Google Cloud~\cite{gke}, it is also open
sourced and usable beyond Google Cloud.  The gVisor infrastructure intercepts most system calls
and reimplements substantial Linux-kernel functionality in user space, reducing exposure to
host-kernel vulnerabilities such as Dirty COW~\cite{dirtycow}.   Its kernel interface is largely
written in Go~\cite{gvisor}, providing memory safety and tighter system-call mediation.

Each protected container receives a logically separate gVisor kernel, comprising the \emph{Sentry}, which implements system-call emulation and core kernel logic, and the \emph{Gofer}, which handles file I/O and communicates with the Sentry via the Plan 9 Filesystem Protocol (9P)~\cite{pike1990plan,gvisor}. %
This design isolates per-container kernel state, preventing direct observation of other containers' system-call behavior or kernel metadata. Prior work shows that gVisor significantly reduces the host-kernel attack surface, although at the cost of syscall overhead~\cite{Anjali2020firecrackerandgVisor}.

The gVisor sandbox supports multiple backends for redirecting execution and
system calls into Sentry. %
We evaluate two platforms
most relevant to current deployments: \emph{systrap}, which uses Linux
\texttt{seccomp} and \texttt{KVM}, which uses
hardware virtualization to improve isolation and address-space switching without
booting a guest kernel.  In both cases, the Sentry remains the application-facing kernel.

This distinction is important because, although gVisor mediates system calls,
it does not necessarily create private copies of host-backed files. Sandboxes
launched from the same container image may access shared
image-layer objects, \emph{exposing hardware timing side channels}. Thus,
even when interfaces such as \texttt{mincore} are unavailable or uninformative,
page-cache state may still be inferred through \emph{hardware-based timing side channels}.

\subsection{KVM}

The Kernel-based Virtual Machine (KVM)~\cite{linuxKVM} is Linux's primary
hardware-assisted virtualization infrastructure. KVM allows the Linux host
kernel to act as a hypervisor by using processor virtualization extensions such
as Intel VT-x or AMD-V. Each virtual machine runs its own guest operating system
with separate kernel state, address spaces, page tables, and virtual CPUs, \emph{exposing various hardware-acceleration timing channels}.
Privileged CPU execution and memory-management virtualization are handled by
KVM in the host kernel, while user-space components such as the Quick Emulator
(QEMU)~\cite{qemu} provide VM orchestration, virtual devices, and disk-image
management.  This architecture provides a stronger isolation boundary than conventional
containers: a process inside the guest cannot directly inspect host-kernel state
or invoke host-level page-cache interfaces on host files. However, VM isolation
does not necessarily eliminate storage-layer hardware sharing. A guest virtual disk is
often backed by a host-side \emph{hardware cached} file, such as a raw disk image or a QEMU
Copy-On-Write version 2 (QCOW2) image. Guest reads and writes are translated by
QEMU into host I/O operations on that backing object. Therefore, when the disk
backend uses the page cache, which resides in the DRAM, file-backed state below the VM boundary can
still introduce %
SCA timing dependencies.%

\subsubsection{QEMU Copy-On-Write (QCOW2) Layers}
\label{sec:bg-qcow2}

QCOW2 is a sparse disk-image format commonly used by QEMU. Instead of allocating
the entire virtual disk eagerly, a QCOW2 image allocates host storage in
clusters as guest blocks are written. The format maintains metadata that maps
guest-visible disk offsets to host-file clusters. If a guest block has not been
allocated in the current image, the read may be satisfied from a backing image,
if one exists, or otherwise from implicit zero-filled data. This backing-file mechanism allows QCOW2 images to form copy-on-write chains.
A read-only base image can contain the common operating-system installation,
while each VM receives a private writable overlay image. Initially, the overlay
contains only metadata and modified clusters. Reads of unchanged guest blocks
are redirected to the shared backing image, whereas writes allocate new clusters
in the overlay image. This is similar in purpose to container image layering:
unchanged data can be shared, while modifications remain private. For page-cache timing leakage, the important point is that QCOW2 sharing occurs through
host-side disk-image files. If two VMs read unchanged data that resolves to the
same host-backed base image, and if QEMU accesses that image through buffered
host I/O, then the corresponding image-file pages may be hardware-cached in the host page
cache. A read by one VM can therefore affect the timing of a later read by
another VM that reaches the same backing-file offset.

\subsubsection{QEMU/KVM Disk Cache Modes}
\label{sec:bg-qemu-cache}

QEMU exposes several disk cache modes that control how host storage is accessed
for a virtual disk. The relevant distinction for this paper is whether guest
block I/O is allowed to use the host page cache. With the default
\texttt{cache=writeback} mode, QEMU uses buffered host I/O and may report writes
as complete once the data has reached the host page cache. In this mode, reads and writes to a host-backed disk
image can populate host page-cache entries. The \texttt{cache=none} mode changes this behavior by enabling direct-I/O-style
access to the backing file. In this configuration, guest data I/O 
bypasses the host page cache (mediating hardware leakage). This distinction is central to our evaluation. If two VMs access shared or
related host-backed disk-image contents under \texttt{cache=writeback}, one VM's
read can warm a host page-cache entry that later affects another VM's timing.
Under \texttt{cache=none}, this interaction is weakened because the same guest
disk access no longer relies on the host page cache in the same way. Thus,
QEMU/KVM provides strong CPU and kernel isolation, but its page-cache leakage
behavior still depends on the disk-image format and cache hardware mode used beneath the
VM boundary.

\subsection{Kata Containers}
Kata Containers~\cite{KataContainers} are a container runtime that provide VM-based
isolation while preserving the container execution model. Instead of running a
container process directly on the host kernel, Kata launches a lightweight
virtual machine, or microVM, and runs the container workload inside that guest, mediated by its fast/slow hardware paths.
The guest contains its own Linux kernel and a Kata agent that coordinates with
the host-side Kata runtime. Kata can be configured with different virtual
machine monitors, including QEMU, Cloud Hypervisor, and Firecracker. This design gives each sandbox a separate kernel and
address space, providing stronger isolation than conventional process-based
containers. Kata introduces an additional storage question that does not exist in ordinary
Linux containers: the container root filesystem must be made visible inside the
guest VM. The root filesystem is first prepared on the host by a container
snapshotter, such as an OverlayFS-based snapshotter or the containerd devmapper
snapshotter. Kata must then expose that prepared filesystem to the guest (utilizing its local hardware resources). In
practice, it can be done using a shared-filesystem path (virtio-fs) or a
block-device path.
\subsubsection{Shared-Filesystem Path: virtio-fs}
\label{sec:bg-kata-virtiofs}

In the shared-filesystem path, \texttt{containerd}’s snapshotter prepares the container root filesystem on the host, and Kata exports it to the guest through \texttt{virtio-fs}. With the OverlayFS snapshotter, shared read-only image layers and each container’s writable layer form a host-side merged view whose file and metadata operations are handled by \texttt{virtiofsd}. Guest accesses can therefore resolve to common host-backed OverlayFS objects, allowing Kata sandboxes created from the same image to share host page-cache state for unchanged layers. Thus, the VM boundary does not eliminate the host-cacheable storage surface.
We evaluate this path for Kata with QEMU and Cloud Hypervisor. Firecracker is evaluated only with block-device-backed root filesystems because Firecracker-based Kata deployments typically use the \texttt{devmapper} snapshotter rather than %
\texttt{virtio-fs}.%

\subsubsection{Block-Device Path}
\label{sec:bg-kata-block}
In the block-device path, the container root filesystem is presented to the Kata guest as a virtual disk rather than exported as a host directory. The guest accesses it through \texttt{virtio-blk} or \texttt{virtio-scsi} and mounts it as the container root filesystem. This path commonly uses the \texttt{devmapper} snapshotter, which represents image and writable layers as thin-provisioned snapshots in a device-mapper thin pool. Unmodified blocks may be shared, while writes allocate private blocks. Thus, sharing occurs below the filesystem level rather than through host-visible file paths.

File operations are resolved inside the guest. Reads may be served from the guest page cache; on a miss, the guest filesystem issues block I/O through the virtual device to the virtual machine monitor and its configured backend. Unlike \texttt{virtio-fs}, this path does not expose individual files, such as \texttt{/bin/sh} or \texttt{/usr/sbin/mysqld}, as host filesystem objects: the guest sees a disk, while the host sees block I/O.

Cache behavior therefore depends not on the block-device abstraction alone, but on the entire storage stack: the guest page cache and filesystem, virtual block device, hypervisor backend, device-mapper thin pool, and physical storage. Their joint configuration determines the strength of side-channel signals.

In particular, a \texttt{devmapper} thin pool can be backed in different ways. A
loop-backed thin pool ultimately stores its data in ordinary host files attached
through loop devices, whereas a real block-device-backed thin pool stores its data
on physical block storage through devices such as LVM logical volumes. The
configurations expose different host-side objects to the storage stack. Hypervisor
cache modes, including buffered and direct-I/O, influence whether guest block I/O uses
the host page cache.
As such, the block-device path should be understood as a family of storage
configurations rather than a single caching behavior. Though it removes the
direct host-file-level sharing interface of \texttt{virtio-fs}, host caching may still occur
depending on the thin-pool backend, loop-device
settings, hypervisor cache policy, and underlying storage device.
\smallskip

\paragraph{Configuration parameters.}
 Kata's storage behavior is controlled by %
(1) the container snapshotter, %
for example as an OverlayFS directory tree or as a
\texttt{devmapper} block-backed snapshot (2) %
the storage-exposure mechanism, which determines how that root filesystem is presented
inside the guest. Options such as
\path{disable_block_device_use}, \path{shared_fs}, and
\path{virtio_fs_cache} select the shared filesystem path; options
like \path{block_device_driver} and
\path{block_device_cache_direct} affect the block-device path. We give the
exact configurations used in our experiments in Subsecs.~\ref{sec:exp-measurement-procedure-overlay} and ~\ref{sec:exp-block-device}.%

\subsection{Device Mapper and Thin Provisioning}
\label{sec:bg-devmapper}

Linux Device Mapper is a kernel framework for constructing virtual block
devices from lower-level block devices. It is used by storage systems such as
LVM, dm-crypt, multipath storage, and dm-thin. In this work, the relevant target
is dm-thin, which provides thin-provisioned block devices and efficient
copy-on-write snapshots.

A dm-thin setup consists of a \emph{thin pool} and one or more \emph{thin
devices}. Each thin device appears to upper layers as an ordinary block device
with its own logical block address space. Internally, however, blocks are
allocated from the shared thin pool only when they are written. The thin pool
maintains metadata that maps each thin device's logical blocks to physical
blocks in the underlying storage. The thin pool itself is backed by two block devices: a data device and a
metadata device. The data device stores the actual contents of allocated blocks,
while the metadata device stores the mapping information required to translate logical to physical addresses. %
This separation allows the thin pool to support snapshots efficiently. Multiple thin devices can initially refer to the same physical blocks, and a new physical block is allocated only when one device writes to a shared block.

Container runtimes can use dm-thin through the containerd devmapper snapshotter.
When a container image is unpacked, the snapshotter represents image layers as a
chain of block-level snapshots. Starting a container from the image creates an
additional writable thin-device snapshot derived from the final image snapshot.
If two containers are started from the same image, their writable snapshots can
share unchanged image data at the block layer, while writes are redirected to
private blocks. This mechanism is analogous in purpose to OverlayFS copy-on-write, but it
operates at block granularity rather than filesystem-directory granularity. %
In Kata Containers, dm-thin is relevant because it can provide a block-backed
container root filesystem. The guest VM sees a virtual disk, while the host
manages the underlying storage through the devmapper snapshotter and the thin
pool. The lower storage used for the thin pool may itself be configured in
different ways, for example using loop devices backed by host files or using
logical volumes on a physical block device. These choices determine the storage
objects below the virtual disk and are therefore important when studying which
cache layers participate in serving guest reads.

\subsection{Page Cache Attacks} %
Side channels exploiting the Linux page cache and other kernel-managed data structures have recently gained interest because they allow an adversary to infer victim activity by probing shared kernel state, often without requiring any privileges. Gruss et al.~\cite{pageCache} demonstrated the first practical page cache side channel using the \texttt{mincore} system call, which reveals whether a given page of a file is resident in memory. Their attack repeatedly evicts pages and probes for residency to infer cross-process activity. The Linux kernel subsequently restricted \texttt{mincore} to mitigate this attack by requiring that callers have appropriate privileges or exclusive file ownership.

Boskov et al.~\cite{boskov2022union} show that these mitigations were insufficient in containerized environments. Their Union Buster attack demonstrated that overlay file systems used by Docker~\cite{docker}, Kubernetes~\cite{kubernetes}, and other container platforms unintentionally reintroduce cross-container visibility into page residency. By combining overlayfs semantics with \texttt{mincore}, they built a covert channel across container boundaries and validated it in a commercial cloud environment---the Microsoft Azure Kubernetes service. Beyond page cache attacks, broader research has explored kernel data structure side channels. Maar et al.~\cite{maar2025kernelsnitch} introduced KernelSnitch, showing that timing differences in system calls involving futexes and timers can leak internal kernel-state properties. These results generalize the observation that kernel-managed structures---such as queues, reference counters, and cache metadata---can be externally inferred through microarchitectural and syscall-level timing.

Our work builds on this line of research but focuses on a different question:
whether page-cache timing signals persist when workloads are separated by
stronger isolation mechanisms in shared-hardware, including sandboxed containers and virtual
machines. Unlike attacks that depend on explicit page-residency interfaces, our
measurement primitive uses only unprivileged access timing. We therefore study
settings where \texttt{mincore()} is unavailable, restricted, or uninformative,
and ask whether storage paths such as OverlayFS, \texttt{virtio-fs},
host-backed VM images, and block-device-backed snapshots still expose observable host-cache effects.

\section{Threat Model}
\label{sec:attack}
We study an \emph{OS-mediated microarchitectural timing side channel} in which an attacker infers whether filesystem-backed memory pages have recently been accessed by a victim through timing observations of the shared page cache. Although the page cache is managed by the operating system, the leakage fundamentally originates from shared architectural resources in the underlying memory hierarchy. Similar to conventional cache-timing attacks, the attacker exploits timing differences introduced by shared performance-optimization mechanisms rather than software logic itself. We consider a local, unprivileged attacker colocated with a victim workload on the same physical host (see, \eg~\cite{collocation2009,UpdatedCollocation2015}). The attacker and victim may execute in different isolation environments, including separate Docker containers, separate gVisor sandboxes, separate Kata Containers, or separate QEMU/KVM virtual machines. The attacker does not possess root privileges, cannot modify the host kernel, hypervisor, container runtime, or hardware configuration, and cannot directly inspect the victim's memory or process state.

The attack assumes that the attacker and victim ultimately access storage objects that share page-cache state beneath their isolation boundary in the same hardware. In shared-filesystem deployments, this naturally occurs when workloads access common container image layers, shared libraries, or other filesystem-backed resources and clearly shared hardware cache and memory hierarchy. In VM-based deployments, guest storage operations may ultimately resolve to shared or related host-backed disk-image contents. The attacker does not require direct interaction with the victim, but must be able to repeatedly access candidate pages whose cache residency may be influenced by the victim's activity. We further assume that the attacker can obtain sufficiently cold probe states
either by inducing eviction through unprivileged memory pressure or by relying
on natural cache churn; privileged cache-flushing mechanisms, when used in our
evaluation, serve only to establish controlled experimental baselines and are
not part of the attacker's capabilities. Unlike prior page-cache attacks that rely on explicit page-cache introspection interfaces such as \texttt{mincore}, we assume that \emph{such interfaces are unavailable, virtualized, or intentionally rendered uninformative}. Instead, the attacker infers the shared page-cache state solely through fine-grained timing measurements obtained from repeated page accesses using an unprivileged timing source such as the x86 \texttt{rdtsc} assembler processor instruction. %

\section{Evaluation}
\label{sec:eval}
We evaluate whether page-cache residency can be inferred across container,
sandbox, and virtual-machine isolation boundaries using timing-only
measurements. The evaluation is organized around four experiments, described below.
We validate the practical significance of our approach through a simplified sender-receiver
communication primitive.%

\textbf{In our first experiment}, we measure whether a cross-boundary timing signal exists across different
isolation mechanisms and storage paths. This experiment compares Docker,
gVisor, Kata Containers, and QEMU/KVM under configurations that expose different
host-side storage surfaces. We evaluate shared-filesystem and host-file-backed
paths, such as OverlayFS, \texttt{virtio-fs}, and host-backed VM images, and
compare them with block-device-backed Kata configurations based on
\texttt{\texttt{devmapper}}. This experiment tests the central claim that leakage depends
not only on the isolation boundary, but also on the storage backend and host
cache policy. \textbf{Second}, we repeat the same timing experiment across three different physical hardware platforms to show how the signal depends on the underlying hardware.
 \textbf{Third}, we study how the timing signal behaves under controlled host-side cache pressure. A natural concern is that background activity on a busy host may hide the signal by adding noise. We therefore introduce competing file-backed cache
activity and measure how the cold and primed timing distributions change as
cache pressure increases. \textbf{Fourth}, we evaluate whether the timing signal can expose semantically meaningful
application behavior. As a case study, we deploy WordPress backed by MySQL and
test whether page-cache observations over selected MySQL-related pages can
distinguish coarse-grained user and administrator actions.

\subsection{Experimental Testbeds}
\label{sec:testbeds}
\paragraph{Testbed overview.}
We use one primary testbed for most of the evaluation and separate testbeds
for two focused experiments. Unless stated otherwise, all results are obtained
on the primary testbed. The communication-primitive validation uses a separate
communication testbed, while the hardware-platform sensitivity experiment
compares three physical hosts. Table~\ref{tab:testbed-usage} summarizes which
testbed is used in each experiment.

\paragraph{Primary testbed:}
Intel Core i5-10500T host with 32~GB RAM (limited to improve eviction) and a 512~GB SSD, running Ubuntu 25.10 (kernel 6.17.0-14-generic). Container tests used gVisor with \texttt{runsc} release-20260112.0, and Kata Containers release 3.26.0. For a hardware-virtualized baseline, we evaluated QEMU/KVM VMs running Ubuntu 22.04.5 LTS, each with 2~GB RAM and VirtIO- backed QCOW storage, running on the same host CPU with KVM acceleration.%
\smallskip

\paragraph{Communication  testbed:}
Intel Core i7-7700 host
with 64~GB RAM and a 2~TB HDD, running Debian GNU/Linux 13. The gVisor
experiments used \texttt{runsc} release-20250820.0.

\paragraph{Cross-hardware platforms:}
All hosts used the same software stack: Linux Mint 22.1,
based on Ubuntu 24.04 LTS, Linux kernel
\texttt{6.8.0-51-generic}, and gVisor \texttt{runsc}
release-\texttt{20260511.0}. Their hardware configurations are summarized in
Tab.~\ref{tab:hardware-platforms}.

\begin{table}[!htb]
\centering
\small
\setlength{\tabcolsep}{4pt}
\renewcommand{\arraystretch}{1.08}
\caption{\textbf{Hardware platforms used in the cross-hardware comparison.}}
\label{tab:hardware-platforms}

\begin{tabularx}{\linewidth}{
    @{}
    c
    >{\raggedright\arraybackslash}l
    c
    >{\raggedright\arraybackslash}X
    @{}
}
\toprule
Host & Processor & Memory & Storage \\
\midrule
H1 & Intel i5-8400 @ 2.80\,GHz
   & 16\,GB DDR4-2133 & Seagate ST1000DM010, 1\,TB SATA HDD  \\%HDD \\

H2 & Intel i7-9700 @ 3.00\,GHz
   & 32\,GB DDR4-2667 & SK hynix PC601, 1\,TB NVMe SSD \\

H3 & Intel i5-10500T @ 2.30\,GHz
   & 32\,GB DDR4-2667 &  SK hynix PC711, 512\,GB NVMe SSD \\
\bottomrule
\end{tabularx}
\end{table}

\begin{table}[t]
\centering
\small
\caption{\textbf{Testbed usage across the evaluation.}}
\label{tab:testbed-usage}
\begin{tabular}{ll}
\toprule
Experiments & Testbeds \\
\midrule
Cross-environment; Cache-pressure;  timing signal; & Primary testbed \\
\qquad WordPress/MySQL &\\
Communication-primitive validation & Communication testbed \\
Hardware-platform sensitivity & Hosts H1--H3 \\
\bottomrule
\end{tabular}
\end{table}

\subsection{Experiment 1: Cross-Environment Timing Signal}
\label{sec:exp_timing_signal}

\paragraph{Goal.}
To determine whether page-cache residency can be detected through timing across different isolation
mechanisms. In particular, we ask whether a page accessed by one platform instance creates a measurable latency shift when later probed
by another instance that shares the same host-backed file state indicating fast hardware mediation.%
\smallskip

\paragraph{Communication Primitive:}
To validate that the observed timing separation is sufficient to support reliable information transfer, we construct a minimal sender--receiver communication primitive using the shared page cache. The sender encodes one bit by priming one of two filesystem-backed pages into the shared page cache. The receiver subsequently probes both pages using the timing primitive described in Sec.~\ref{sec:attack}. By comparing the measured access latencies, the receiver infers which page remains resident and therefore recovers the %
bit. %
Unlike traditional page-cache attacks based on explicit page-residency interfaces such as \texttt{mincore}, this communication primitive relies solely on timing observations by %
repeated page accesses using \texttt{rdtsc} x86 (inline) assembly instruction. %

Both the sender and receiver are implemented in C using only standard POSIX file I/O. We deliberately avoid \texttt{mmap} and pointer dereferencing to minimize architectural effects such as hardware prefetching, speculative memory accesses, and TLB warm-up. The sender primes selected filesystem-backed pages into the shared page cache through file reads, the receiver probes the same pages using repeated \texttt{rdtsc} reads.

\begin{figure*}[!b]
    \centering
    \includegraphics[width=0.90\linewidth, height=4.9cm, trim=0 0.3cm 0 0.3cm, clip]{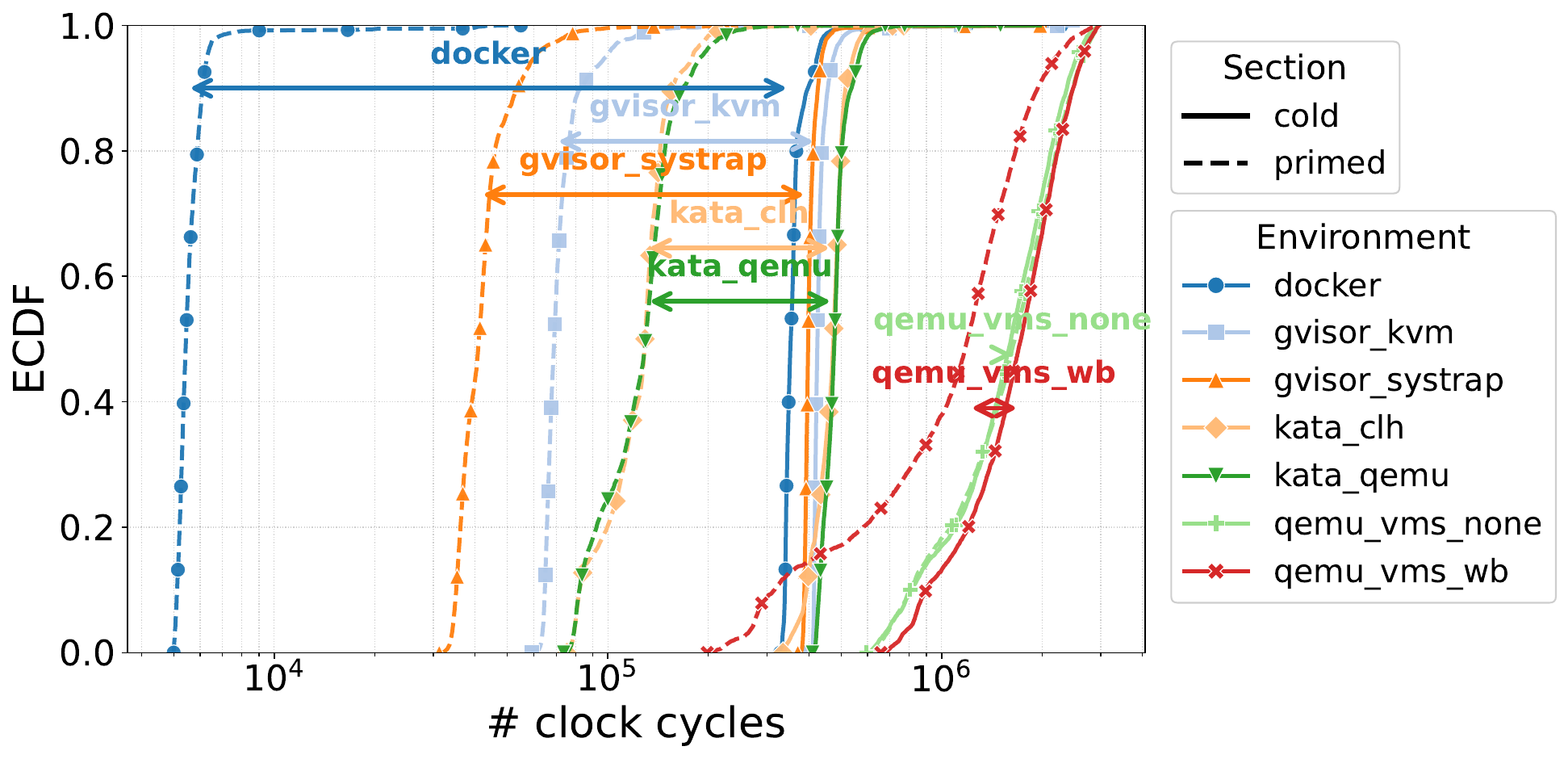}
    \caption{\textbf{Cross-environment timing signal.}  ECDF of receiver page-probing latency in CPU cycles across isolation environments. Solid lines show the cold phase; dashed lines show the primed phase after a separate isolated instance accesses the target page.}
    \label{fig:env_ecdf}
\end{figure*}

\subsubsection{Host/Shared-Filesystem Path: OverlayFS and virtio-fs}
\label{sec:exp-measurement-procedure-overlay}

\paragraph{Setup:} We first evaluate isolation environments whose storage path ultimately exposes
host-side filesystem objects. For native Docker, the container root filesystem
is provided through OverlayFS. For gVisor, the runtime similarly operates on a
host-prepared container root filesystem, while file operations are mediated
through gVisor's Sentry/Gofer architecture. For Kata Containers in the shared-filesystem configuration, the host-side root filesystem is exported into the guest using \texttt{virtio-fs}. For QEMU/KVM, we evaluate VM disk images under different host cache policies. Although these systems differ in their isolation mechanisms, they share an important property: the pages accessed by the isolated workload ultimately
interact with host-side backing state. Table~\ref{tab:host-shared-envs} summarizes the evaluated environments.
\begin{table}[!htb]
    \centering
    \small
    \setlength{\tabcolsep}{4pt}
    \caption{Isolation environments evaluated in the host/shared-filesystem path.}
    \label{tab:host-shared-envs}
    \begin{tabularx}{\columnwidth}{@{}lX@{}}
        \toprule
        \textbf{Environment} & \textbf{Configuration} \\
        \midrule
        \texttt{docker} &
        Docker with the regular \texttt{runc} runtime. \\

        \texttt{gvisor\_kvm} &
        gVisor using \texttt{runsc} with the \texttt{KVM} backend. \\

        \texttt{gvisor\_systrap} &
        gVisor using \texttt{runsc} with the \texttt{systrap} backend. \\

        \texttt{kata\_clh} &
        Kata Containers with Cloud Hypervisor using shared
        \texttt{virtio-fs} storage path. \\

        \texttt{kata\_qemu} &
        Kata Containers with QEMU using the shared
        \texttt{virtio-fs} storage path. \\

        \texttt{qemu\_vms\_none} &
        QEMU/KVM VMs using a host-backed disk image with
        \texttt{cache=none}. \\

        \texttt{qemu\_vms\_wb} &
        QEMU/KVM VMs using a host-backed disk image with
        \texttt{cache=writeback}. \\
        \bottomrule
    \end{tabularx}
\end{table}

Each environment is measured
under two controlled phases: 
\begin{itemize}
  \setlength{\itemsep}{0pt}
  \setlength{\parsep}{0pt}
    \item \textbf{Cold phase:} the receiver probes the target page after cache reset and before any cooperating platform instance accesses the page.
    \item \textbf{Primed phase:} a separate platform instance first accesses the same target page, and the receiver then repeats the same timing measurement. For each environment and phase, we collect 10{,}000 timing samples.
\end{itemize}
To reduce distortion from scheduling and I/O outliers, we apply robust
trimming and retain only measurements within the 1st--99th percentile
range.

For the communication-primitive experiments, the receiver classifies each probe using an empirically calibrated timing threshold determined independently for every platform from the corresponding cold and primed timing distributions. This calibration accounts for differences in processor microarchitecture, storage devices, operating-system implementation, virtualization mechanisms, and background system activity.

\paragraph{Results:} 
Fig.~\ref{fig:env_ecdf} presents the empirical cumulative distribution
function (ECDF) of page-probing latency across all evaluated
environments. For any cycle threshold $\tau$, the ECDF gives the
fraction of probes that completed within $\tau$ cycles. Therefore, a
left-shifted curve indicates lower access latency. The solid curves correspond to the cold phase, while the dashed curves correspond to the primed phase. Each color represents a different
isolation environment. A clear horizontal separation between the solid and dashed curves indicates that priming by one platform instance changes the receiver's timing distribution, which is the timing signal needed for page-cache side-channel inference.

We observe a consistent timing shift for Docker, both gVisor backends,
both Kata configurations, and QEMU/KVM with \texttt{cache=writeback}.
In these configurations, the primed phase is generally faster than the
cold phase, indicating that cross-instance access leaves a measurable
cache footprint. In contrast, for QEMU/KVM with \texttt{cache=none},
the cold and primed distributions largely overlap, and the timing shift disappears. This suggests
that the attack depends not only on virtualization boundaries, but also
on the storage path and host cache policy.

\subsubsection{Block-Device Path: Virtual Disk Backends}
\label{sec:exp-block-device}
\paragraph{Setup.}

We next evaluate the same timing question on a block-device-backed storage
path. In this configuration, Kata Containers does not expose the container
root filesystem through a shared host filesystem such as \texttt{virtio-fs}.
Instead, the container image and writable container state are represented by the
\texttt{containerd} \texttt{\texttt{devmapper}} snapshotter using a \texttt{dm-thin} pool. Each Kata
container is then booted with its root filesystem attached to the guest as a
virtual block device.
This path removes the direct shared-filesystem interface between the guest and
the host. A read issued by a process inside the container is first resolved by
the guest kernel, guest page cache, and guest filesystem. On a guest cache miss,
the request is translated into block I/O and sent through the virtual block
device to the virtual machine monitor, which then resolves the request through
the host-side \texttt{devmapper} storage stack. We evaluate two \texttt{dm-thin} backend variants. In the first variant, the \texttt{dm-thin} pool
is backed by sparse host files attached through loop devices, similar to the
common \texttt{containerd} \texttt{devmapper} quickstart configuration. In this case, the loop
backing files are ordinary host filesystem objects, and their contents may be
cached in the host page cache. We therefore use the loop direct-I/O setting as a
cache-control switch: with loop direct I/O disabled, the loop-backing files may use
the host page cache; with loop direct I/O enabled, this host-file page-cache
surface is largely bypassed. In the second variant, the \texttt{dm-thin} pool is backed by a dedicated physical block
device managed through LVM. The device is divided into two logical volumes: one
used as the \texttt{dm-thin} data device and one used as the metadata device, it %
avoids loop-backing files, and therefore removes the
ordinary host-file object that exists underneath the loop-backed pool.

For QEMU and Cloud Hypervisor, we additionally vary the virtual block-device
cache configuration exposed by Kata. In particular, we compare configurations in
which the virtual machine monitor may use buffered block I/O with configurations
that use direct-I/O-style access. Firecracker exposes a more limited
block-device interface in this setup, so we evaluate it under its default
block-device path rather than applying the same explicit VMM-level cache toggle. The measurement procedure follows the trial-based structure described in
Subsec.~\ref{sec:exp-measurement-procedure-overlay} for 200 iterations. Each trial resets the relevant
cache state, starts two Kata instances from the same image, and records one
receiver probe. In the cold phase, the receiver reads the target page without a
prior access by the second instance. In the primed phase, the second instance
first reads the same target page.%
\smallskip

\paragraph{Results:} Table~\ref{tab:kata-block-results} summarizes the block-device results.
The loop-backed block-device configuration with loop direct I/O disabled shows a
clear timing signal. Across the evaluated Kata hypervisors, the receiver's
target-page access is consistently faster in the primed phase than in the cold
phase. Thus, even though the two containers execute inside separate Kata virtual
machines, an access by one instance can leave a measurable footprint that is
later observed by the other. However, this signal disappears when loop direct I/O is enabled. In that case,
the loop-backing files no longer provide the same host page-cache surface, and
the cold and primed timing distributions largely overlap. We observe the same
qualitative behavior when replacing the loop-backed pool with the real
block-device-backed \texttt{dm-thin} pool. With the internal block-device backend, the
primed phase does not produce a consistent speedup over the cold phase. 
The reason is that \texttt{dm-thin} shares storage blocks, but not necessarily the corresponding
host page-cache state.
In the loop-backed configuration, the \texttt{dm-thin} data device is
backed by an ordinary host file (\eg the \texttt{\texttt{containerd}} \texttt{devmapper} data
file). When two thin snapshots reference the same unchanged image block, their
reads can converge on the same offset. If loop direct I/O is
disabled, this file offset may be cached in the host page cache. Thus, a read by one
Kata VM can warm a host page-cache entry that is later reused by
another VM, producing the observed timing shift. In the real block-device configuration, this backing-file layer is removed. The
two containers may still share unchanged image blocks through \texttt{dm-thin} metadata,
but this sharing occurs at the block-mapping layer, below the file object that
would normally provide a shared page-cache identity. The host does not simply
treat the shared lower physical block as a single file-backed page-cache entry
that is automatically reused across the two virtual block devices. As a result,
a read by one VM may warm its own guest page cache or lower storage state, but it
does not create the same reusable host file-cache entry for the other VM.

These results show that the timing signal observed in the loop-backed
configuration %
is not merely a consequence of \texttt{dm-thin} block sharing, but rather depends on the presence of
a host-cacheable backing file underneath the thin pool. When this 
page-cache surface is eliminated, either by enabling loop direct I/O or by using a
real block-device backend, the cross-VM timing signal is no longer observed.
Together with the shared-filesystem results, these findings confirm that the
timing signal depends not only by the isolation mechanism, but also on
the underlying storage backend and the cache policy. Shared-filesystem and
loop-backed configurations expose host-cacheable file objects, enabling
cross-instance timing signals, whereas direct-I/O and real block-device
configurations weaken them.

 \begin{table}[!htb]
\centering
\small
\setlength{\tabcolsep}{4.0pt}
\renewcommand{\arraystretch}{1.08}
\caption{\textbf{Kata block-device timing results.}
Mean receiver access latency is reported in thousands of CPU cycles
(kC) over $200$ samples per cell. The ratio is
$R=\bar{T}_{\mathrm{cold}}/\bar{T}_{\mathrm{primed}}$; $R\approx1$
indicates little or no timing separation, while $R>1$ (in \textbf{bold}) means the primed
read is faster. ``Cache'' is the VMM
\texttt{block\_device\_cache\_direct} setting. Firecracker (FC) exposes
no equivalent knob. CLH denotes Cloud Hypervisor.}
\label{tab:kata-block-results}
\begin{tabular}{@{}llllcrrr@{}}
\toprule
Backend & HV & Driver & Mode & Cache
& $\bar{T}_{\mathrm{cold}} (kC)$
& $\bar{T}_{\mathrm{primed}}(kC)$
& $R$ \\
\midrule
Loop \texttt{dm-thin} & QEMU & virtio-blk  & DIO off & on  & $625.8$ & $345.7$ & $\mathbf{1.81}$ \\
Loop \texttt{dm-thin} & QEMU & virtio-blk  & DIO off & off & $624.2$ & $366.6$ & $\mathbf{1.70}$ \\
Loop \texttt{dm-thin} & QEMU & virtio-blk  & DIO on  & on  & $575.9$ & $645.9$ & $0.89$ \\
Loop \texttt{dm-thin} & QEMU & virtio-blk  & DIO on  & off & $556.7$ & $702.3$ & $0.79$ \\
Loop \texttt{dm-thin} & CLH  & virtio-blk  & DIO off & on  & $287.8$ & $101.5$ & $\mathbf{2.84}$ \\
Loop \texttt{dm-thin} & CLH  & virtio-blk  & DIO off & off & $307.7$ & $117.9$ & $\mathbf{2.61}$ \\
Loop \texttt{dm-thin} & CLH  & virtio-blk  & DIO on  & on  & $277.1$ & $284.2$ & $0.97$ \\
Loop \texttt{dm-thin} & CLH  & virtio-blk  & DIO on  & off & $300.4$ & $308.5$ & $0.97$ \\
Loop \texttt{dm-thin} & FC   & virtio-mmio & DIO off & --  & $551.4$ & $308.1$ & $\mathbf{1.79}$ \\
Loop \texttt{dm-thin} & FC   & virtio-mmio & DIO on  & --  & $515.0$ & $497.7$ & $\mathbf{1.03}$ \\
\midrule
Real \texttt{dm-thin} & QEMU & virtio-blk  & block & on  & $453.1$ & $472.4$ & $0.96$ \\
Real \texttt{dm-thin} & QEMU & virtio-blk  & block & off & $422.5$ & $411.5$ & $\mathbf{1.03}$ \\
Real \texttt{dm-thin} & CLH  & virtio-blk  & block & on  & $242.9$ & $246.2$ & $0.99$ \\
Real \texttt{dm-thin} & CLH  & virtio-blk  & block & off & $250.4$ & $282.0$ & $0.89$ \\
Real \texttt{dm-thin} & FC   & virtio-mmio & block & --  & $344.5$ & $351.5$ & $0.98$ \\
\bottomrule
\end{tabular}
\end{table}

\subsubsection{Communication Primitive Validation}

Although the ECDF analysis establishes the existence of a measurable timing signal, it does not directly indicate whether the observed latency separation is sufficient to support reliable communication. We therefore evaluate the communication primitive described above by measuring its bit error rate and achievable throughput across environments.%

\subparagraph{Timing Channel Reliability:} Table~\ref{tab:runtime_comm} summarizes the measured bit error rates across representative runtime environments. The reported $1\!\rightarrow\!0$ and $0\!\rightarrow\!1$ errors correspond to false negatives and false positives when inferring the page-cache state solely from timing observations. The results show that the characterized timing leakage remains highly observable across all evaluated platforms. Docker and gVisor-KVM exhibit overall bit error rates below $0.1\%$, indicating that cached and uncached page accesses remain highly distinguishable despite container or virtualization boundaries. Although gVisor-systrap introduces additional software mediation that increases timing variability, the overall error rate remains below $2\%$. These results indicate that stronger isolation mechanisms primarily reduce the signal-to-noise ratio (SNR) of the timing channel rather than eliminating the underlying shared architectural state responsible for the leakage.
\begin{table}[!hbt]
\centering
\small
\begin{tabular}{l c ccc}
\toprule
&
&
\multicolumn{3}{c}{\textbf{Bit Error Rate (\%)}}\\
\cmidrule(lr){3-5}
\textbf{Runtime} &
\textbf{Bandwidth (bps)} &
\textbf{$1\!\rightarrow\!0$} &
\textbf{$0\!\rightarrow\!1$} &
\textbf{Overall} \\
\midrule
docker-runc    & 477.55 & 0.0000 & 0.0000 & 0.0000 \\
gvisor-kvm     & 357.20 & 0.0637 & 0.0573 & 0.0605 \\
gvisor-systrap & 290.84 & 0.7703 & 3.1877 & 1.9648 \\
\bottomrule
\end{tabular}
\caption{Communication performance. %
The table reports the achieved bandwidth and corresponding bit error rates. A $1\!\rightarrow\!0$ error denotes a primed page classified as unprimed, whereas a $0\!\rightarrow\!1$ error denotes an unprimed page classified as primed. }
\label{tab:runtime_comm}
\end{table}

\subparagraph{Communication Throughput:} To estimate the effective capacity of the timing channel, we measure the average time required to transmit a 1024-bit message. Let $T_s$ denote the sender execution time and $T_r$ the receiver execution time. The effective throughput is computed as $B=\frac{\texttt{message.length}}{\max(T_s,T_r)}$. Table~\ref{tab:runtime_comm} reports the resulting throughput across the evaluated runtimes. 
As expected, additional software mediation introduced by gVisor reduces throughput compared with native Docker execution because each transmitted bit requires additional page accesses and timing measurements. Nevertheless, all evaluated environments provide sufficient channel stability to support reliable communication.

Together with the ECDF measurements presented earlier, these results demonstrate that the observed timing separation is not merely statistically distinguishable, but sufficiently robust to support practical communication across multiple virtualization boundaries. The communication primitive therefore serves as an independent validation that the page-cache microarchitectural timing leakage characterized is practically exploitable.%

\subsection{Experiment 2: Hardware-Platform Sensitivity}
\label{sec:hardware-sensitivity}
\paragraph{Goal:}
We evaluate the timing signal across different physical hardware platforms.

\paragraph{Setup:}
We repeat the same timing experiment on three hosts for 1000 iterations, while keeping
the software stack and experimental configuration fixed. The hardware specifications of the three hosts are summarized under Cross-hardware platforms in Subsec.~\ref{sec:testbeds}.
\smallskip

\begin{figure}[tb]
    \centering
    \includegraphics[width=\linewidth]{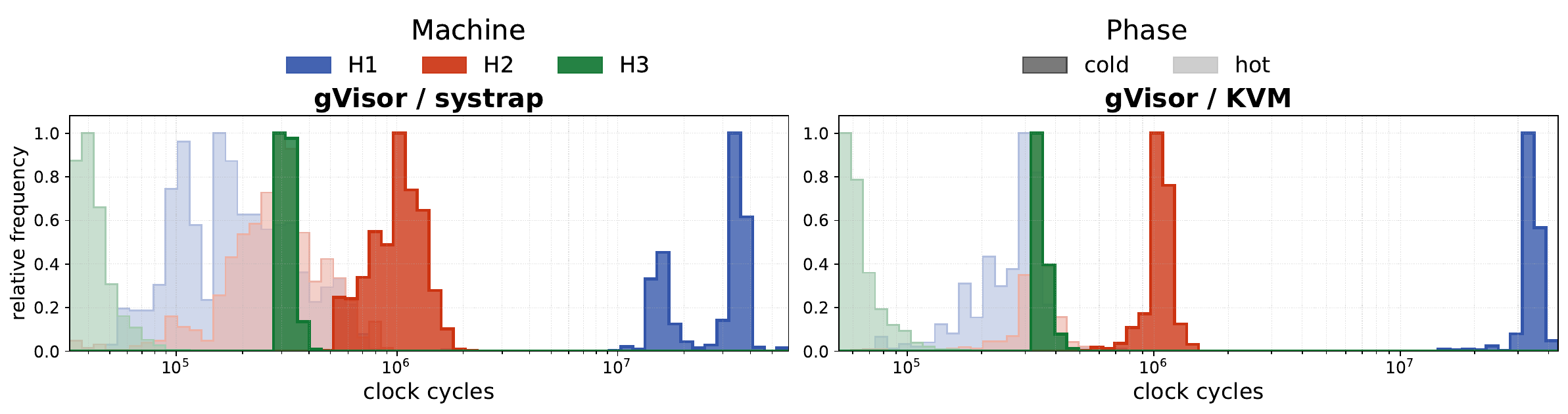}
    \caption{\textbf{Timing distributions across three hardware platforms.}
    Peak-normalized cold and hot access-time distributions for gVisor (systrap
    and KVM) on hosts H1--H3.}
    \label{fig:hardware-platform-comparison}
\end{figure}

\paragraph{Results:}
As shown in Fig.~\ref{fig:hardware-platform-comparison},
the observed timing distributions differ across the three hardware platforms because the complete hardware path—including the processor, memory hierarchy, storage interface, and storage device—shapes both cached and uncached access latency.

\subsection{Experiment 3: Cache-Pressure Timing Signal Amplification}
\label{sec:exp_noise}

\paragraph{Goal:} Experiment~\ref{sec:exp_timing_signal} showed that page-cache residency creates
a measurable timing difference between cold and recently primed accesses across
several isolation mechanisms and storage paths. In this experiment, we evaluate
whether this timing signal survives under a noisier and more realistic host
setting. In particular, we ask whether background file-cache activity hides the
signal by adding noise and evicting the target page, or whether it can amplify
the timing gap between cold and recently primed accesses.

This experiment is motivated by a natural concern: a page-cache timing signal
measured on a quiet laboratory machine may disappear on a busy multi-tenant
host, where unrelated workloads continuously access files and compete for
memory. %
\smallskip

\paragraph{Setup:} We repeat the same two-phase measurement procedure used in
Experiment~\ref{sec:exp_timing_signal}. In the cold phase, the receiver probes a
target page without a recent access by another instance. In the primed phase, a
second instance first accesses the same target page, and the receiver then
measures its own access latency. To increase cache pressure, we limit the host memory to 8~GB and run a
controlled host-side noise workload in parallel with the measurement loop. The
noise workload consists of \(N\) independent worker processes. Each worker first
creates a 128~MB buffer file and then repeatedly reads the entire file in a
tight loop: $\texttt{cat buffer > /dev/null}$. Thus, the nominal file-backed cache footprint of the noise workload is
\(128N\)~MB. The repeated reads create several forms of interference. First, the buffer-file
pages compete with the target page in the Linux page cache. Increasing \(N\)
therefore increases the probability that the target page is reclaimed before the
cold probe. Second, the workers repeatedly stream through memory, creating
memory-bandwidth pressure. Third, the use of \texttt{cat} introduces CPU and
syscall overhead, since each worker repeatedly issues file reads and writes the
data to \texttt{/dev/null}. Finally, when the noise working set exceeds the
available cache, or during the initial population of the buffers, the workload
can also introduce storage-I/O pressure. We therefore view this workload
primarily as a controlled page-cache and memory-pressure generator, with
possible additional storage pressure depending on the available device memory.%

We evaluate Docker, gVisor, and Kata Containers using the following noise
levels: $N \in \{0,30,40,50,60\}$. These correspond to nominal file-backed footprints of 0~GB, 3.84~GB, 5.12~GB,
6.4~GB, and 7.68~GB, respectively. For QEMU/KVM virtual machines, we use a
smaller sweep,
$N \in \{0,30,35,38\}$, because the VM setting has higher baseline memory demand and could not 
run under the larger noise configurations. %
For each environment and noise level, we record the probing latency in CPU clock-cycles for the cold and primed phases. We also compute the amplification ratio, $A(N) = \frac{\overline{T}_{\mathrm{cold}}(N)}{\overline{T}_{\mathrm{primed}}(N)}$,
where \(\overline{T}_{\mathrm{cold}}(N)\) and
\(\overline{T}_{\mathrm{primed}}(N)\) are the mean cycle counts under \(N\)
noise workers. A larger value of \(A(N)\) indicates stronger timing separability
between the two states.

\begin{figure}[tb]
    \centering
    \includegraphics[width=\textwidth, trim=0 0.8cm 0 0.8cm, clip]{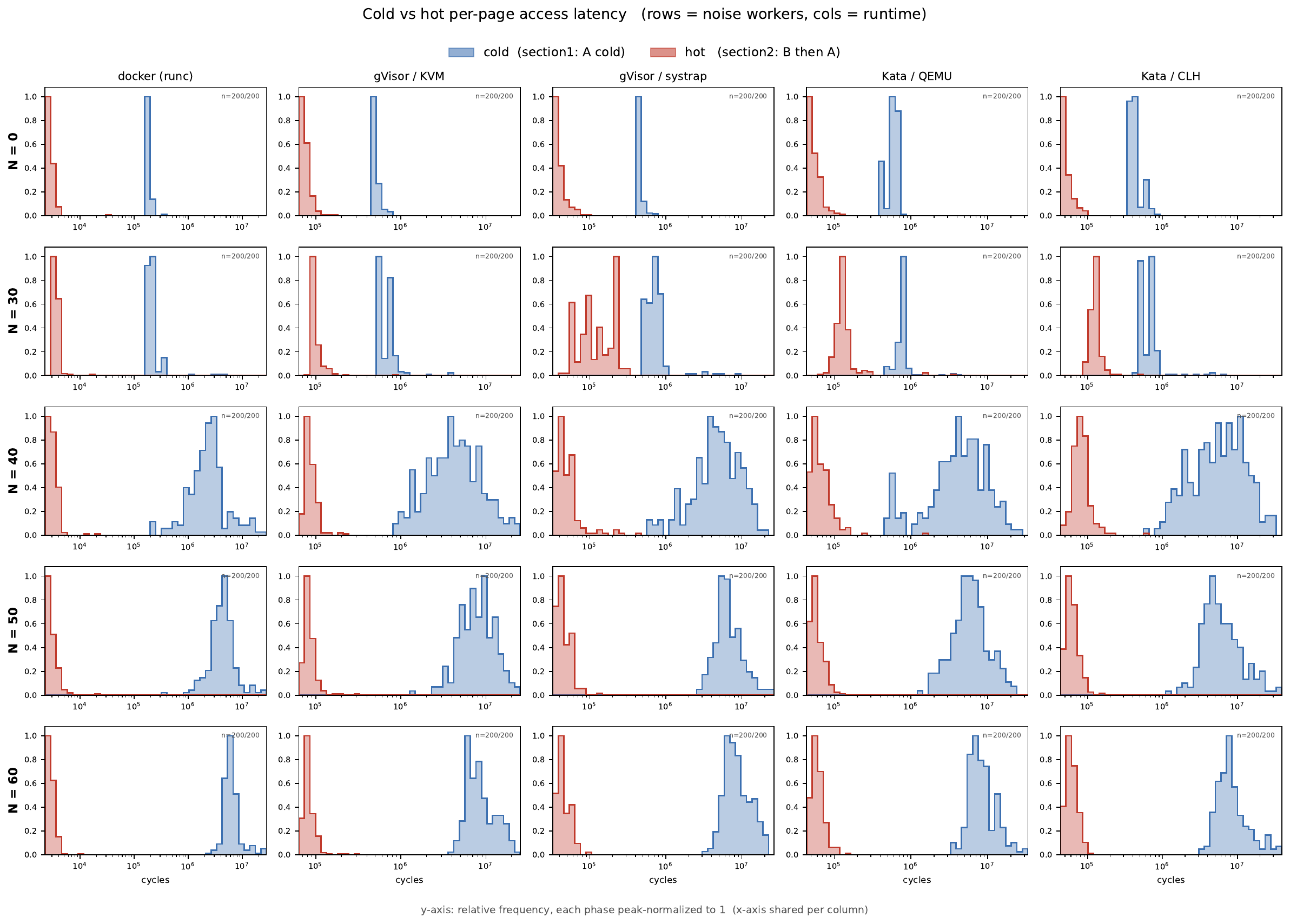}
    \caption{Cold and primed per-page access latency under increasing
    host-side noise for Docker, gVisor, and Kata Containers. Rows correspond to
    the number of noise workers \(N\), and columns correspond to runtimes. The
    x-axis is latency in CPU cycles on a log scale. The y-axis shows relative
    frequency, with each phase peak-normalized to one.}
    \label{fig:noise_env_grid}
\end{figure}

\begin{figure}[tb]
    \centering
    \includegraphics[width=0.88\textwidth, trim=0 0.5cm 0 0.5cm, clip]{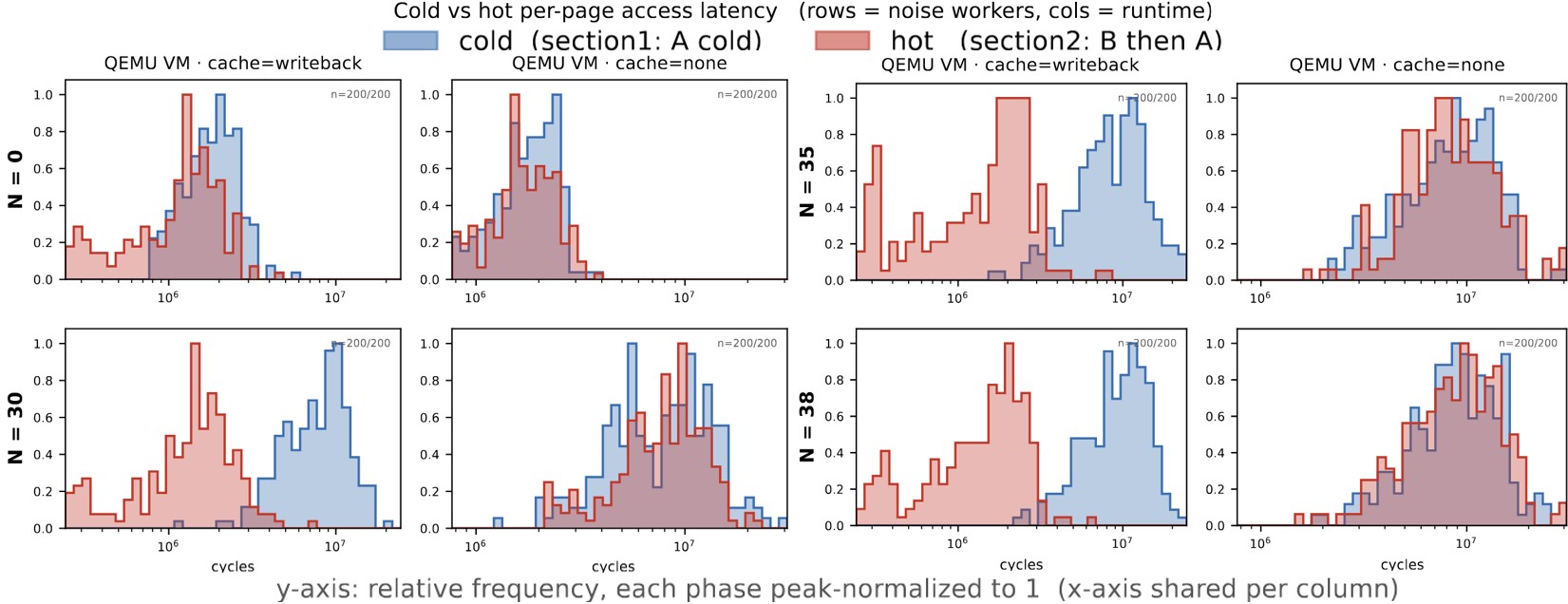}
    \caption{Cold and primed per-page access latency under increasing
    host-side noise for QEMU/KVM virtual machines. The VM experiment uses a
    smaller noise sweep, \(N \in \{0,30,35,38\}\), because the VM setting could
    not reliably run under the larger noise configurations used for the
    container and Kata experiments.}
    \label{fig:qemu_vms_grid}
\end{figure}

\textbf{Results:} 
Fig.~\ref{fig:noise_env_grid} shows the effect of increasing host-side
file-backed cache pressure on Docker, gVisor, and Kata Containers. Across these
environments, the cold phase is affected much more strongly than the primed
phase. As the number of noise workers increases, the cold distribution shifts
rightward by orders of magnitude, while the primed distribution remains
comparatively close to the low-latency region. This behavior shows that the noise workload does not simply destroy the
page-cache timing signal. Instead, it selectively penalizes the cold condition.
In the cold phase, the target page has not been recently accessed by the
cooperating instance. The
receiver's access then follows the slow miss path: the page must be refilled
through the storage stack, and the probing thread or virtual CPU may also
experience queueing and scheduling delay before the access completes. 
In the primed phase, the cooperating instance accesses the target page
immediately before the receiver probes it. This access re-primes the page shortly
before measurement. Although the noise workers continue to churn through
file-backed cache pages, the time window between priming and probing is short
enough that the freshly populated target page usually remains available. The
primed phase therefore remains much less sensitive to increasing \(N\). Fig.~\ref{fig:qemu_vms_grid} shows the same experiment for QEMU/KVM virtual
machines. The
\texttt{cache=writeback} configuration shows the same qualitative amplification
effect: increasing noise makes the cold distribution substantially slower while
the primed distribution remains faster. In contrast, the \texttt{cache=none}
configuration shows no cold/primed separation, as in
Experiment~\ref{sec:exp_timing_signal}.

Consequently, the ratio
\(\overline{T}_{\mathrm{cold}}/\overline{T}_{\mathrm{primed}}\) grows with
increasing noise in configurations where the storage path exposes a relevant
host-cacheable object. The counterintuitive result is an \emph{amplification} of the timing gap, showing this timing SCA path is not a quiet-machine artifact.

\subsection{Experiment 4: Application-Level WordPress/MySQL Side Channel}
\textbf{Goal:} Having established that the timing signal is both observable and sufficiently reliable to support communication, we next investigate whether the same signal can reveal semantically meaningful application behavior. 
Our proof-of-concept (PoC) deployment consists of a WordPress stack running on our testbed. The environment includes four containers:
\textbf{($i$)} \texttt{wp\_app}  - WordPress \textbf{($ii$)} \texttt{wp\_db}  - MySQL~8.4 with the general query log enabled \textbf{($iii$)} \texttt{prober}  - attacker-controlled page-cache probing using either mincore or timing with a configurable cycle threshold to classify probes as cached vs. uncached \textbf{($iv$)} \texttt{evictor} - controlled eviction via memory pressure. To ensure a fast and reliable cache reset, we pin \texttt{evictor} to the native \texttt{runc} runtime in all configurations. We then repeat the same workload while varying the runtime of the remaining containers to evaluate whether leakage persists across stronger isolation boundaries.

\subparagraph{Eviction mechanism:}
 We use two eviction primitives, matched to each configuration's threat model.
  For the native Docker (\texttt{runc}) configuration, which serves as an
  \emph{oracle upper bound}, we assume host privileges and flush the cache with
  the \texttt{drop\_caches} interface (writing \texttt{/proc/sys/vm/drop\_caches}),
  which unconditionally evicts \emph{all} clean page-cache entries and thus yields
  the cleanest possible cold state---appropriate for the \texttt{mincore} oracle,
  which scores every page of the binary.
  For the sandboxed gVisor configurations, which model a \emph{realistic,
  unprivileged} attacker that cannot invoke privileged kernel interfaces, the
  \texttt{evictor} container induces eviction purely through memory pressure: it
  allocates and repeatedly accesses a large buffer of size \texttt{EVICT\_SIZE},
 causing the host kernel's LRU-based
  reclaim to evict previously cached \texttt{mysqld} pages from the host page
  cache without special privileges. Each trial therefore follows the sequence: $\textit{stop}
  \rightarrow
  \textit{evict}
  \rightarrow
  \textit{start}
  \rightarrow
  \textit{act}
  \rightarrow
  \textit{probe}$. The eviction step is performed before the monitored action to establish a
consistent cold-cache baseline for each trial.

\subparagraph{Monitored actions:}  We select six representative WordPress actions spanning both public and administrative behavior:
\texttt{logout},
\texttt{login\_success},
\texttt{password\_reset},
\texttt{view\_homepage},
\texttt{publish\_post},
and \texttt{add\_user}.
These actions induce distinct sequences of database queries and filesystem activity, which in turn create different page-cache residency patterns for the database server binary and its dependent libraries. Note that our testbed does not configure outbound email delivery (\eg SMTP/relay), so the password-reset action is measured up to the point where WordPress attempts to send the reset email, which triggers a UI error although the server-side processing and database activity still occur.

\subparagraph{Template construction:}  For each action $a$, we construct a probabilistic template by repeating the following procedure $N$ = 30 times:
\textbf{($1$)} evict memory \textbf{($2$)} execute $a$ \textbf{($3$)} probe a fixed set of $M$ file-backed pages of \texttt{/usr/sbin/mysqld}, yielding a binary observation vector $\mathbf{x}\in\{0,1\}^M$ indicating page residency. We then estimate an action-specific probability vector, $\mathbf{p}_a = (p_{a,1}, p_{a,2}, \dots, p_{a,M})$, where $p_{a,i}$ is the empirical probability that page $i$ is resident after executing action $a$.
\smallskip

\subparagraph{Independence assumption (prod.-Bernoulli):}
We approximate the whole bitmap's joint distribution $X = (X_1, \ldots, X_M)$ as independent bits:
\[P_a(X = x) \approx \prod_{i=1}^{M} \left(p_i^{(a)}\right)^{x_i}\left(1 - p_i^{(a)}\right)^{1-x_i}.\]

This is a \emph{Bernoulli product model}.
To quantify separability between actions, we compute the Jensen--Shannon (JS) distance  between templates~\cite{jenson_shanon_distance}: where
\[
\mathrm{JS}(\mathbf{p}_a,\mathbf{p}_b) =\sqrt{\tfrac12 D_{\mathrm{KL}}(\mathbf{p}_a\|\mathbf{m}) +\tfrac12 D_{\mathrm{KL}}(\mathbf{p}_b\|\mathbf{m})} =\sqrt{\tfrac12\sum_{i=1}^{M} [d(p_{a,i},m_i)+d(p_{b,i},m_i)]},\] with $d(x,y)
:=
D_{\mathrm{KL}}\!\bigl(\mathrm{Ber}(x)\,\|\,\mathrm{Ber}(y)\bigr)$, 
where $\mathrm{Ber}(p)$ is the Bernoulli distribution with parameter $p$, and $D_{\mathrm{KL}}(X\|Y)$ is the Kullback--Leibler divergence between distributions $X$ and $Y$. Larger JS distances indicate stronger statistical separation between actions.

\paragraph{Classification experiment (Proof-of-Concept):} We evaluate whether these templates enable action inference using a maximum-likelihood classifier. In each trial, we evict memory, execute one action (unknown to the classifier), probe the monitored pages to obtain $\mathbf{x}$, and score each candidate action $a$ via the Bernoulli log-likelihood $\log L(a \mid \mathbf{x})
=$ 
$\sum_{i=1}^{M}
\Big(
x_i \log p_{a,i} + (1-x_i)\log(1-p_{a,i})
\Big)$. We rank actions by score and report Top-1 and Top-2 accuracy over $N=50$ repetitions per action.
\smallskip

\paragraph{Results:} We evaluate application-level side channel that infers high-level WordPress actions from page-cache effects in the MySQL server binary, with %
configurations:\\
\noindent  \textbf{(1)} Docker with \texttt{runc} using \texttt{mincore}-based probing and privileged \texttt{drop\_caches} eviction (oracle upper bound). \textbf{(2)} gVisor systrap using timing %
probing with smart-stride page selection and unprivileged memory-pressure eviction. \textbf{(3)} gVisor KVM using timing-based probing with smart-stride page selection and unprivileged memory-pressure eviction.

\begin{figure}[tb]
\centering
\includegraphics[width=0.81\linewidth]{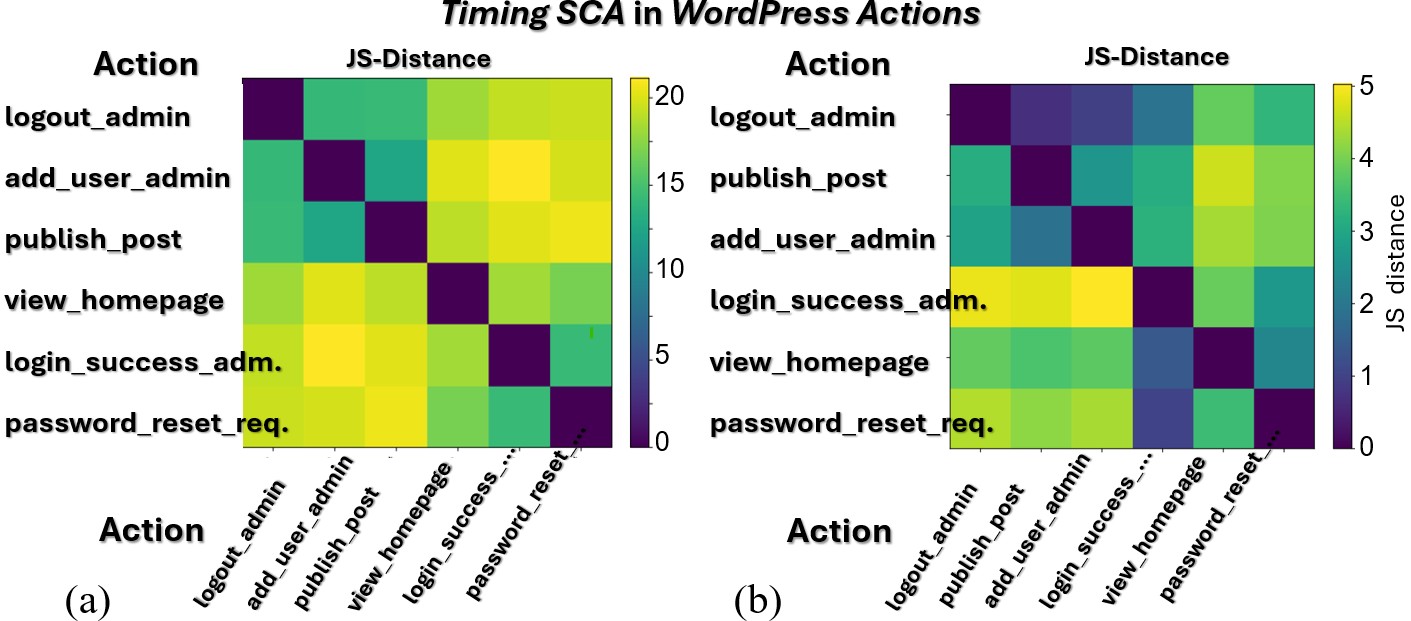}
\caption{  (a) Jensen--Shannon distance matrix of the templates for Docker attack - using all pages of /usr/sbin/mysqld file. (b) Jensen--Shannon distance matrix of the templates for gVisor attack - using the chosen subset of /usr/sbin/mysqld file pages.
}
\label{fig_wordpress_a_b}
\end{figure}

\textbf{Docker (runc) with \texttt{mincore}:}
In the native container configuration, where \texttt{mincore} is available, we directly observe page residency after each action execution. Fig.~\ref{fig_wordpress_a_b}(a) shows the resulting JS distance matrix between action templates, Pairs of actions with small JS distance are less distinguishable and thus more likely to be confused by the classifier.  We then apply a maximum-likelihood classifier over the per-action models; the corresponding per-action classification accuracies are reported in Table~\ref{tab:sidechannel_summary}.

\textbf{gVisor (Systrap and KVM) with timing + smart stride:}
Under gVisor, system mediation renders \texttt{mincore} unavailable; thus, we rely only on timing-based probing. In addition, host-level prefetching may fetch up to 128\,KB (32 pages) per access. To reduce this effect, pages that are at least 32 pages apart (smart stride) are probed. When templating actions, we also compute a mask, denoted \texttt{mask\_before} of persistently resident pages. Concretely, after performing eviction but \emph{before} executing any monitored action, we probe the target file (\texttt{/usr/sbin/mysqld}) and mark repeatedly resident pages. These pages capture stable background cache state and are excluded from the probing set.
  
\textbf{Discriminative page selection:} For each action, we rank candidate pages by the per-page Bernoulli Kullback--Leibler
  divergence between that action's template and the mean of the remaining actions, and retain
  pages that are (i) reliably resident after the action ($p_i \ge \texttt{p\_min}$),
  (ii) discriminative ($\mathrm{KL}_i \ge \texttt{kl\_min}$), and (iii) at least
  \texttt{stride} pages apart and outside \texttt{mask\_before}. Each action is thus probed
  and scored over its own selected page subset. Because different actions select different
  pages, the resulting JS distance matrix (Fig.~\ref{fig_wordpress_a_b}(b)) is
  asymmetric: row $i$ measures how separable each other action is from action $i$ over
  \emph{action~$i$'s} probed pages, the geometry that predicts the
  classifier's confusions.

  \begin{table}[!htb]
  \centering
  \small
  \setlength{\tabcolsep}{4pt}
  \begin{tabular}{lcc|cc|cc}
  \toprule
  & \multicolumn{2}{c|}{\textbf{Docker (runc)}} &
  \multicolumn{2}{c|}{\textbf{gVisor systrap}} &
  \multicolumn{2}{c}{\textbf{gVisor KVM}} \\
  \cmidrule(r){2-3}
  \cmidrule(r){4-5}
  \cmidrule(l){6-7}
  \textbf{Action} &
  \topOne & \topTwo &
  \topOne & \topTwo &
  \topOne & \topTwo \\
  \midrule
  add\_user\_admin         & 100\% & 100\% &  96\% &  98\% &   0\% &   8\% \\
  login\_success\_admin    & 100\% & 100\% & 100\% & 100\% &   8\% &  58\% \\
  logout\_admin            & 100\% & 100\% & 100\% & 100\% & 100\% & 100\% \\
  password\_reset\_request & 100\% & 100\% & 100\% & 100\% &  12\% &  70\% \\
  publish\_post            & 100\% & 100\% & 100\% & 100\% & 100\% & 100\% \\
  view\_homepage           &  98\% & 100\% &  90\% &  96\% &  78\% &  88\% \\
  \midrule
  \textbf{Mean}            & \textbf{99.7\%} & \textbf{100\%}
                            & \textbf{97.7\%} & \textbf{99.0\%}
                            & \textbf{49.7\%} & \textbf{70.7\%} \\
  \bottomrule
  \end{tabular}
  \caption{Per-action Top-1/-2 classification accuracy across the evaluated execution environments (N=50 trials/action). The last row reports the
  mean accuracy.}%
  \label{tab:sidechannel_summary}
  \end{table}

\textbf{Classification Results:}
Table~\ref{tab:sidechannel_summary}  shows that Docker and gVisor systrap achieve near-perfect classification, with mean \topOne accuracies of 99.7\% and 97.7\%, respectively. Note that the Docker baseline additionally assumes host privileges
  (\texttt{drop\_caches} eviction) and therefore
  represents an \emph{upper bound} on leakage. The gVisor configurations rely only
  on unprivileged primitives (timing probing and memory-pressure eviction) and
  thus reflect a realistic attacker operating across the isolation boundary. Under gVisor KVM, the mean \topOne accuracy decreases to 49.7\%, indicating that hardware virtualization introduces additional noise and reduces template separability. Nevertheless, the accuracy remains above the 16.7\%  \topOne random-guessing baseline, demonstrating that action-dependent timing leakage persists across the isolation boundary.

\section{Conclusion}
\label{sec:conclusions}

We have shown that page-cache timing leakage persists across modern container, sandbox, and virtual-machine isolation boundaries whenever the storage path exposes shared host-cacheable state. Although the page cache is managed by the operating system, the resulting leakage constitutes an OS-mediated microarchitectural timing side channel whose observability is jointly determined by the processor microarchitecture, memory hierarchy, storage architecture, virtualization mechanism, and host cache policy. Across Docker, gVisor, Kata Containers, and QEMU/KVM, stronger software isolation reshapes the observable timing signal but does not inherently eliminate the underlying
leakage.

Our evaluation further demonstrates that leakage depends not merely on the existence of shared storage objects, but on how those objects are reached through the storage stack. Shared file-backed objects, including those based on OverlayFS layers, virtio-fs exports, and loop-backed block devices, preserve observable timing state, whereas direct-I/O-style access and dedicated block-device-backed storage largely attenuate or eliminate the signal; moreover background activity does not necessarily conceal, and in some cases accentuates, the side channel. Our WordPress/MySQL case study demonstrates that the channel extends beyond synthetic benchmarks and can reveal coarse-grained application behavior.

Overall, our findings place page-cache attacks within the broader class of architectural and microarchitectural timing side channels. Effective mitigation thus requires coordinated hardware, operating-system, virtualization, and storage-stack support to provide robust timing isolation rather than relying solely on stronger software isolation boundaries. Ultimately, page-cache leakage should be viewed not as a purely operating-system artifact, but as an OS-mediated microarchitectural timing side channel arising from shared hardware-supported performance state.

\section*{Acknowledgments}
We thank Fredrik Wilke and Alexander Choi for identifying the gVisor page-cache side channel, which inspired this work, and Fredrik Wilke and Yahav Kadar for their contributions to a preliminary precursor of this work.

This work was supported in part by the National Science Foundation's Industry–University Cooperative Research Centers (IUCRC) Program, Division of Computer and Network Systems, under Award No. 1939098. We also thank the IUCRC Industry Advisory Board for its thoughtful feedback. Any opinions, findings, conclusions, or recommendations expressed in this material are those of the authors and do not necessarily reflect the views of the National Science Foundation. Itamar Levi and Alon Abudraham were partly founded by the Israel Science Foundation (ISF) grant 2569/21 and 3276/25.

\newpage

\bibliographystyle{abbrv} 
\bibliography{refs}

\begin{thebibliography}{10}

\bibitem{Anjali2020firecrackerandgVisor}
Anjali, T.~Caraza-Harter, and M.~M. Swift.
\newblock Blending containers and virtual machines: a study of firecracker and gvisor.
\newblock In {\em Proceedings of the 16th ACM SIGPLAN/SIGOPS International Conference on Virtual Execution Environments}, VEE '20, page 101–113, New York, NY, USA, 2020. Association for Computing Machinery.

\bibitem{boskov2022union}
N.~Boskov, N.~Radami, T.~Tiwari, and A.~Trachtenberg.
\newblock Union buster: A cross-container covert-channel exploiting union mounting.
\newblock In {\em International Symposium on Cyber Security, Cryptology, and Machine Learning}, pages 300--317. Springer, 2022.

\bibitem{burns2016borg}
B.~Burns, B.~Grant, D.~Oppenheimer, E.~Brewer, and J.~Wilkes.
\newblock Borg, omega, and kubernetes.
\newblock {\em Commun. ACM}, 59(5):50–57, Apr. 2016.

\bibitem{docker}
{Docker Inc.}
\newblock {Docker: Accelerated Container Application Development}, 2025.

\bibitem{linuxKVM}
S.~French.
\newblock {Linux KVM Documentation}, 2025.

\bibitem{linuxPageCache}
S.~French.
\newblock {Linux Page Cache Documentation}, 2025.

\bibitem{gke}
{Google Cloud}.
\newblock {Google Kubernetes Engine Documentation}, 2025.

\bibitem{pageCache}
D.~Gruss, E.~Kraft, T.~Tiwari, M.~Schwarz, A.~Trachtenberg, J.~Hennessey, A.~Ionescu, and A.~Fogh.
\newblock Page cache attacks.
\newblock In {\em Proceedings of the 2019 ACM SIGSAC Conference on Computer and Communications Security}, CCS '19, page 167–180, New York, NY, USA, 2019. Association for Computing Machinery.

\bibitem{jenson_shanon_distance}
J.~Lin.
\newblock Divergence measures based on the shannon entropy.
\newblock {\em IEEE Transactions on Information Theory}, 37(1):145--151, 1991.

\bibitem{maar2025kernelsnitch}
L.~Maar, J.~Juffinger, T.~Steinbauer, D.~Gruss, and S.~Mangard.
\newblock Kernelsnitch: Side-channel attacks on kernel data structures.
\newblock In {\em Network and Distributed System Security Symposium 2025: NDSS 2025}, 2025.

\bibitem{dirtycow}
P.~Oester.
\newblock {CVE-2016-5195: Dirty COW - Privilege Escalation Linux Kernel}, 2016.

\bibitem{osvik2006cache}
D.~A. Osvik, A.~Shamir, and E.~Tromer.
\newblock Cache attacks and countermeasures: The case of {AES}.
\newblock In {\em Topics in Cryptology -- {CT-RSA} 2006}, volume 3860 of {\em Lecture Notes in Computer Science}, pages 1--20. Springer, 2006.

\bibitem{pike1990plan}
R.~Pike, D.~Presotto, K.~Thompson, H.~Trickey, et~al.
\newblock Plan 9 from bell labs.
\newblock In {\em Proceedings of the summer 1990 UKUUG Conference}, pages 1--9. London, UK, 1990.

\bibitem{collocation2009}
T.~Ristenpart, E.~Tromer, H.~Shacham, and S.~Savage.
\newblock Hey, you, get off of my cloud: exploring information leakage in third-party compute clouds.
\newblock In {\em Proceedings of the 16th ACM Conference on Computer and Communications Security}, CCS '09, page 199–212, New York, NY, USA, 2009. Association for Computing Machinery.

\bibitem{gvisor}
{The gVisor Authors}.
\newblock {gVisor}, 2018.

\bibitem{KataContainers}
{The Kata Containers Authors}.
\newblock {Kata Containers}, 2018.

\bibitem{kubernetes}
{The Kubernetes Authors}.
\newblock {Kubernetes}, 2025.

\bibitem{qemu}
{The QEMU Project Developers}.
\newblock {QEMU Documentation}, 2025.

\bibitem{UpdatedCollocation2015}
V.~Varadarajan, Y.~Zhang, T.~Ristenpart, and M.~Swift.
\newblock A placement vulnerability study in {Multi-Tenant} public clouds.
\newblock In {\em 24th USENIX Security Symposium (USENIX Security 15)}, pages 913--928, Washington, D.C., Aug. 2015. USENIX Association.

\bibitem{borg}
A.~Verma, L.~Pedrosa, M.~Korupolu, D.~Oppenheimer, E.~Tune, and J.~Wilkes.
\newblock Large-scale cluster management at google with borg.
\newblock In {\em Proceedings of the Tenth European Conference on Computer Systems}, EuroSys '15, New York, NY, USA, 2015. Association for Computing Machinery.

\bibitem{yarom2014flush}
Y.~Yarom and K.~Falkner.
\newblock $\{$FLUSH+ RELOAD$\}$: A high resolution, low noise, l3 cache $\{$Side-Channel$\}$ attack.
\newblock In {\em 23rd USENIX security symposium (USENIX security 14)}, pages 719--732, 2014.

\end{thebibliography}
\newpage

 \appendix
  \section{Kata Containers Configuration Details}
  \label{app:kata-config}

  This appendix summarizes the Kata Containers storage options used in our
  experiments. These options are included for reproducibility. The conceptual
  difference between the shared-filesystem and block-device paths is described in
  Subsecs.~\ref{sec:bg-kata-virtiofs} and .~\ref{sec:bg-kata-block}.

  \subsection{Snapshotter and Guest-Exposure Layers}
  \label{app:kata-snapshotter-exposure}

  Kata storage configuration involves two distinct layers. The first layer is the
  container snapshotter, which determines how the container root filesystem is
  materialized on the host. For example, the OverlayFS snapshotter produces a
  host-side merged directory tree, while the \path{devmapper}
  snapshotter produces block-backed snapshots.

  The second layer is the Kata guest-exposure mechanism, which determines how the
  prepared root filesystem is made visible inside the guest VM. A host-side
  directory tree can be exported through a shared filesystem such as
  \path{virtio-fs}. A block-backed root filesystem can instead be attached to
  the guest as a virtual block device, for example using \path{virtio-blk} or
  \path{virtio-scsi}.

  \subsection{Shared-Filesystem Configuration}
  \label{app:kata-virtiofs-config}

  For the shared-filesystem path, we use an OverlayFS-based snapshotter and expose
  the resulting host-side root filesystem through \path{virtio-fs}. A
  representative configuration is:

  \begin{verbatim}
  disable_block_device_use = true
  shared_fs = "virtio-fs"
  virtio_fs_cache = "auto"
  \end{verbatim}

  The option \path{disable_block_device_use=true} prevents Kata from passing
  the root filesystem as a block device. Instead, Kata exposes the mounted
  host-side root filesystem using the configured shared filesystem. The option
  \path{shared_fs="virtio-fs"} selects \path{virtio-fs} as the host--guest
  filesystem transport.

  The option \path{virtio_fs_cache} controls caching behavior in the
  \path{virtio-fs} path. The standard Kata Containers configuration installed
  with the runtime uses \path{auto} as the default value, and we therefore use
  \path{auto} as our baseline shared-filesystem setting. When stricter control
  is needed, \path{never} can be used to remove guest-side \path{virtio-fs}
  caching.

  \subsection{Block-Device Configuration}
  \label{app:kata-block-config}

  For the block-device path, we use the containerd \path{devmapper} snapshotter
  to provide a block-backed container root filesystem. Kata is then allowed to
  attach this root filesystem to the guest as a virtual block device. A
  representative configuration is:

  \begin{verbatim}
  disable_block_device_use = false
  block_device_driver = "virtio-blk"
  \end{verbatim}

  The option \path{disable_block_device_use=false} allows Kata to use a
  block-backed root filesystem when the snapshotter provides one. This option does
  not create a block device by itself; it only permits Kata to pass one into the
  guest.

  The option \path{block_device_driver} selects the guest-visible block-device
  interface. The Kata default for QEMU is \path{virtio-scsi}; to keep the
  guest-visible interface identical across QEMU and Cloud Hypervisor, we set
  \path{virtio-blk} for both. Firecracker is an exception and is discussed in
  Subsec.~\ref{app:kata-hypervisor-storage}.

  \subsection{Block-Device Cache Controls}
  \label{app:kata-block-cache-config}

  For block-backed root filesystems, Kata exposes cache-related options that
  control how the virtual machine monitor opens the block backend:

  \begin{verbatim}
  block_device_cache_set = true
  block_device_cache_direct = false
  block_device_cache_noflush = false
  \end{verbatim}

  The option \path{block_device_cache_set} controls whether Kata explicitly
  sets the block-device cache policy. When it is set to \path{false}, the
  hypervisor default is used. For controlled experiments, we set it to
  \path{true} when the hypervisor supports the option.

  The option \path{block_device_cache_direct} controls whether the virtual
  machine monitor uses direct-I/O-style access for the block backend. When it is
  set to \path{false}, guest block I/O may use buffered host I/O, so the host
  page cache backing the block device can be populated and shared. When it is set
  to \path{true}, the backend is opened using direct-I/O-style semantics where
  supported, bypassing the host page cache. This option is the primary
  host-cache lever we sweep in our block-device experiments.

  The option \path{block_device_cache_noflush} controls whether guest flush
  requests may be ignored. We keep this option set to \path{false}, so that
  guest flush requests are honored.

  \subsection{Host-Side Block Backend: Real Device vs.\ Loop-Backed Pool}
  \label{app:kata-loop-backend}

  The Kata cache options of Subsec.~\ref{app:kata-block-cache-config} control how
  the \emph{guest} virtual machine monitor opens its block backend. Independently,
  the host determines what that block backend physically \emph{is}. We evaluate two
  containerd \path{devmapper} thin-pool backends:

  \begin{itemize}
    \item \textbf{Real block device.} The thin-pool is provisioned on a dedicated
          partition of the internal NVMe device. There is no host-side backing
          file: the pool's data and metadata are plain logical volumes on the
          device.
    \item \textbf{Loop-backed pool.} The thin-pool data and metadata devices are
          sparse regular files on the host filesystem, attached as block devices
          through Linux loop devices (\path{losetup}).
  \end{itemize}

  The loop backend exposes an additional host-side lever, the loop device's
  direct-I/O flag (\path{losetup -{}-direct-io}, denoted \path{LOOP_DIO}):

  \begin{itemize}
    \item \path{LOOP_DIO=off} (default): the loop device performs buffered I/O
          against its backing file, so the backing file's contents are cached in
          the \emph{host} page cache. Because that cache is keyed by the backing
          file, it is shared by any VM whose block device is served from the same
          file. This is the host-side analogue of
          \path{block_device_cache_direct=false}.
    \item \path{LOOP_DIO=on}: the loop device opens its backing file with
          \path{O_DIRECT}, bypassing the host page cache, so no shared
          file-backed cache is populated.
  \end{itemize}

  This distinction is central to our results: a cross-VM timing signal appears only
  when a shared host-side cache exists over a common backing object. The
  loop-backed pool with \path{LOOP_DIO=off} provides exactly such an object (the
  shared backing file), whereas the real block device has no backing file and the
  host caches it per block-device inode, leaving no cross-VM file cache to share.
  Because \path{LOOP_DIO} is a host loop-device flag rather than a Kata option,
  it applies uniformly to all three hypervisors, including Firecracker, even though
  Firecracker exposes no \path{block_device_cache_direct} control of its own.

  \subsection{Hypervisor-Specific Storage Support}
  \label{app:kata-hypervisor-storage}

  The three hypervisors do not expose the same set of storage options. QEMU and
  Cloud Hypervisor support both the shared-filesystem and block-device paths, and
  both honor the block-device cache controls of
  Subsec.~\ref{app:kata-block-cache-config}. Firecracker is more restricted, and
  these restrictions directly shape which experiments are possible:

  \begin{itemize}
    \item \textbf{No shared-filesystem path.} The Kata Firecracker integration
          does not support \path{virtio-fs}; it requires a block-based root
          filesystem provided by the \path{devmapper} snapshotter. Consequently
          \path{disable_block_device_use=true} is not usable with
          Firecracker, and Firecracker is always evaluated on the block-device
          path.
    \item \textbf{MMIO transport.} It does not expose the same \path{virtio-blk} driver
selection used by QEMU/CLH in our Kata configuration.
    \item \textbf{No direct-I/O cache control.} Firecracker does not implement an
          equivalent of \path{block_device_cache_direct}. We therefore cannot
          sweep the direct-I/O lever for Firecracker.
  \end{itemize}

  \begin{table}[!htb]
  \centering
  \caption{Kata storage paths and block-device options as used in our evaluation.}
  \label{tab:kata-hypervisor-storage}
  \begin{tabular}{@{}llll@{}}
  \toprule
  Hypervisor & Shared-FS path & Block driver (this work) & \path{cache_direct} \\
  \midrule
  QEMU &
  \path{virtio-fs} &
  \path{virtio-blk} &
  supported \\

  Cloud Hypervisor &
  \path{virtio-fs} &
  \path{virtio-blk} &
  supported \\

  Firecracker &
  not supported &
  \path{virtio-mmio} &
  not supported \\
  \bottomrule
  \end{tabular}
  \end{table}

  \subsection{Configuration of Each Evaluated Run}
  \label{app:kata-run-config}

Tab.~\ref{tab:kata-run-config} gives the run-level configuration for the
Kata block-device experiment described in
Subsec.~\ref{sec:exp-block-device} and summarized in
Tab.~\ref{tab:kata-block-results}. The evaluation table reports the measured
cold and primed access latencies; this appendix table records the exact storage
levers used to produce each row.
 The column
\path{cache_direct} corresponds to the \path{Cache} column in
Tab.~\ref{tab:kata-block-results} and denotes Kata's
\path{block_device_cache_direct} setting. The column \path{LOOP_DIO}
denotes whether direct I/O was enabled on the host loop devices backing the
dm-thin pool. A dash (---) marks a lever that does not apply: real block-device
runs do not use loop devices, and in our Firecracker configuration we do not
vary Kata's \path{block_device_cache_direct} option.

  \begin{table}[!htb]
  \centering
  \small
  \caption{Host and guest storage levers for each evaluated run
  (200 samples per run).}
  \label{tab:kata-run-config}
  \begin{tabular}{@{}lllcc@{}}
  \toprule
  Run identifier & Hyperv. & Pool backend
  & \path{cache_direct} & \path{LOOP_DIO} \\
  \midrule
  \multicolumn{5}{@{}l}{\emph{Real block device (internal NVMe)}}\\
  \path{qemu_internal_cdon}  & QEMU & block & on  & --- \\
  \path{qemu_internal_cdoff} & QEMU & block & off & --- \\
  \path{clh_internal_cdon}   & CLH  & block & on  & --- \\
  \path{clh_internal_cdoff}  & CLH  & block & off & --- \\
  \path{kata_fc_internal}    & FC   & block & --- & --- \\
  \midrule
  \multicolumn{5}{@{}l}{\emph{Loop-backed thin-pool}}\\
  \path{qemu_loop_diooff_cdon}  & QEMU & loop & on  & off \\
  \path{qemu_loop_diooff_cdoff} & QEMU & loop & off & off \\
  \path{qemu_loop_dioon_cdon}   & QEMU & loop & on  & on  \\
  \path{qemu_loop_dioon_cdoff}  & QEMU & loop & off & on  \\
  \path{clh_loop_diooff_cdon}   & CLH  & loop & on  & off \\
  \path{clh_loop_diooff_cdoff}  & CLH  & loop & off & off \\
  \path{clh_loop_dioon_cdon}    & CLH  & loop & on  & on  \\
  \path{clh_loop_dioon_cdoff}   & CLH  & loop & off & on  \\
  \path{kata_fc_loop_diooff}    & FC   & loop & --- & off \\
  \path{kata_fc_loop_dioon}     & FC   & loop & --- & on  \\
  \bottomrule
  \end{tabular}
  \end{table}

\end{document}